%% file: main.tex
\pdfoutput=1
\documentclass[12pt,a4paper]{article}
\usepackage{ifthen} % for conditional statements
\newboolean{pdflatex}
\setboolean{pdflatex}{true} % False for eps figures 

\newboolean{articletitles}
\setboolean{articletitles}{true} % False removes titles in references

\newboolean{uprightparticles}
\setboolean{uprightparticles}{true} %True for upright particle symbols

\newboolean{completeplots}
\setboolean{completeplots}{false}

\def\paperauthors{LHCb collaboration} % Leave as is for PAPER, CONF and FIGURE
\def\paperasciititle{Observation of the~decay Lb->Chic1PPi-} % Set ASCII title here !! MAKE sure it's only ASCII characters !! 
\def\papertitle{Observation of the~decay \LbChiconePPi} % Latex formatted title
\def\paperkeywords{{High Energy Physics}, {LHCb}} % Comma separated list
\def\papercopyright{\the\year\ CERN for the benefit of the LHCb collaboration} % new since 9/Apr/2018
\def\paperlicence{CC BY 4.0 licence}
\def\paperlicenceurl{https://creativecommons.org/licenses/by/4.0/}

\input{preamble}
\input{local_symbols} % Add in the additional commands
\usepackage{graphpap}
\usepackage{comment}

\makeatletter
\g@addto@macro\bfseries{\boldmath}
\makeatother

\begin{document}

%% Title
\renewcommand{\thefootnote}{\fnsymbol{footnote}}
\setcounter{footnote}{1}
\input{title-LHCb-PAPER}

%%
\renewcommand{\thefootnote}{\arabic{footnote}}
\setcounter{footnote}{0}

%% Table of Content
%% \tableofcontents
\cleardoublepage

%% Main text
\pagestyle{plain} % restore page numbers for the main text
\setcounter{page}{1}
\pagenumbering{arabic}
%% \linenumbers

%% Main body
\input{introduction}
\input{detector}
\input{selection}
\input{signals}
\input{systematics}
\input{results}

%% Acknowledgements
\input{acknowledgements}

%% \clearpage 
%% \input{supplementary-app}

%% References
\addcontentsline{toc}{section}{References}
%\setboolean{inbibliography}{true}
\bibliographystyle{LHCb}
\bibliography{main,standard,LHCb-PAPER,LHCb-CONF,LHCb-DP,LHCb-TDR}

 %% Authorship
\newpage
\input{Authorship_LHCb-PAPER-2021-003}

\end{document}

%% file: preamble.tex
\usepackage[top=1in, bottom=1.25in, left=1in, right=1in]{geometry}

\usepackage{microtype}
\usepackage[mathlines]{lineno}  % for line numbering during review
\usepackage{xspace} % To avoid problems with missing or double spaces after
\usepackage{caption} %these three command get the figure and table captions automatically small

\usepackage{graphicx}  % to include figures (can also use other packages)
\usepackage{color}
\usepackage{colortbl}
\graphicspath{{./figs/}} % Make Latex search fig subdir for figures
\usepackage{amsmath} % Adds a large collection of math symbols
\usepackage{amssymb}
\usepackage{amsfonts}
\usepackage{upgreek} % Adds in support for greek letters in roman typeset

\newcommand*\patchAmsMathEnvironmentForLineno[1]{%
\expandafter\let\csname old#1\expandafter\endcsname\csname #1\endcsname
\expandafter\let\csname oldend#1\expandafter\endcsname\csname
end#1\endcsname
 \renewenvironment{#1}%
   {\linenomath\csname old#1\endcsname}%
   {\csname oldend#1\endcsname\endlinenomath}%
}
\newcommand*\patchBothAmsMathEnvironmentsForLineno[1]{%
  \patchAmsMathEnvironmentForLineno{#1}%
  \patchAmsMathEnvironmentForLineno{#1*}%
}
\AtBeginDocument{%
\patchBothAmsMathEnvironmentsForLineno{equation}%
\patchBothAmsMathEnvironmentsForLineno{align}%
\patchBothAmsMathEnvironmentsForLineno{flalign}%
\patchBothAmsMathEnvironmentsForLineno{alignat}%
\patchBothAmsMathEnvironmentsForLineno{gather}%
\patchBothAmsMathEnvironmentsForLineno{multline}%
\patchBothAmsMathEnvironmentsForLineno{eqnarray}%
}

\usepackage{hyperxmp}

\usepackage[pdftex,
            pdfauthor={\paperauthors},
            pdftitle={\paperasciititle},
            pdfkeywords={\paperkeywords},
            pdfcopyright={Copyright (C) \papercopyright},
            pdflicenseurl={\paperlicenceurl}]{hyperref}
\usepackage[colorinlistoftodos,textsize=scriptsize]{todonotes}

\usepackage[bottom,flushmargin,hang,multiple]{footmisc}

\usepackage[all]{hypcap} % Internal hyperlinks to floats.

\input{lhcb-symbols-def} % Add in the predefined LHCb symbols

\usepackage{cite} % Allows for ranges in citations
\usepackage{mciteplus}

%% file: lhcb-symbols-def.tex
\usepackage{xspace} 
\usepackage{upgreek}

\def\lhcb   {\mbox{LHCb}\xspace}

\def\MagUp {\mbox{\em Mag\kern -0.05em Up}\xspace}

\ifthenelse{\boolean{uprightparticles}}%
{
 
 \def\Pgamma      {\ensuremath{\upgamma}\xspace}

 \def\Peta        {\ensuremath{\upeta}\xspace}
 \def\Ptheta      {\ensuremath{\uptheta}\xspace}

 \def\Pmu         {\ensuremath{\upmu}\xspace}

 \def\Ppi         {\ensuremath{\uppi}\xspace}

 \def\Pphi        {\ensuremath{\upphi}\xspace}                 
                  
 \def\Pchi        {\ensuremath{\upchi}\xspace}                 
 \def\Ppsi        {\ensuremath{\uppsi}\xspace}

 \def\PDelta      {\ensuremath{\Delta}\xspace}                 
 \def\PXi         {\ensuremath{\Xi}\xspace}                 
 \def\PLambda     {\ensuremath{\Lambda}\xspace}                 
 \def\PSigma      {\ensuremath{\Sigma}\xspace}                 
 \def\POmega      {\ensuremath{\Omega}\xspace}                 
 \def\PUpsilon    {\ensuremath{\Upsilon}\xspace}

 \def\PB      {\ensuremath{\mathrm{B}}\xspace}                 
                  
 \def\PD      {\ensuremath{\mathrm{D}}\xspace}

 \def\PJ      {\ensuremath{\mathrm{J}}\xspace}                 
 \def\PK      {\ensuremath{\mathrm{K}}\xspace}

 \def\PS      {\ensuremath{\mathrm{S}}\xspace}

 \def\PX      {\ensuremath{\mathrm{X}}\xspace}

 \def\Pb      {\ensuremath{\mathrm{b}}\xspace}                 
 \def\Pc      {\ensuremath{\mathrm{c}}\xspace}

 \def\Pi      {\ensuremath{\mathrm{i}}\xspace}

 \def\Pp      {\ensuremath{\mathrm{p}}\xspace}

 \def\Ps      {\ensuremath{\mathrm{s}}\xspace}

 \def\thebaroffset{0.0em}
}
{
 
 \def\Pgamma      {\ensuremath{\gamma}\xspace}

 \def\Peta        {\ensuremath{\eta}\xspace}                 
 \def\Ptheta      {\ensuremath{\theta}\xspace}

 \def\Pmu         {\ensuremath{\mu}\xspace}

 \def\Ppi         {\ensuremath{\pi}\xspace}

 \def\Pphi        {\ensuremath{\phi}\xspace}                 
                  
 \def\Pchi        {\ensuremath{\chi}\xspace}                 
 \def\Ppsi        {\ensuremath{\psi}\xspace}                 
                  
 \mathchardef\PDelta="7101
 \mathchardef\PXi="7104
 \mathchardef\PLambda="7103
 \mathchardef\PSigma="7106
 \mathchardef\POmega="710A
 \mathchardef\PUpsilon="7107
                  
 \def\PB      {\ensuremath{B}\xspace}                 
                  
 \def\PD      {\ensuremath{D}\xspace}

 \def\PJ      {\ensuremath{J}\xspace}                 
 \def\PK      {\ensuremath{K}\xspace}

 \def\PS      {\ensuremath{S}\xspace}

 \def\PX      {\ensuremath{X}\xspace}

 \def\Pb      {\ensuremath{b}\xspace}                 
 \def\Pc      {\ensuremath{c}\xspace}

 \def\Pi      {\ensuremath{i}\xspace}

 \def\Pp      {\ensuremath{p}\xspace}

 \def\Ps      {\ensuremath{s}\xspace}

 \def\thebaroffset{0.18em}
}
\newcommand{\offsetoverline}[2][\thebaroffset]{\kern #1\overline{\kern -#1 #2}}%

\makeatletter
\ifcase \@ptsize \relax% 10pt
  \newcommand{\miniscule}{\@setfontsize\miniscule{4}{5}}% \tiny: 5/6
\or% 11pt
  \newcommand{\miniscule}{\@setfontsize\miniscule{5}{6}}% \tiny: 6/7
\or% 12pt
  \newcommand{\miniscule}{\@setfontsize\miniscule{5}{6}}% \tiny: 6/7
\fi
\makeatother

\DeclareRobustCommand{\optbar}[1]{\shortstack{{\miniscule (\rule[.5ex]{1.25em}{.18mm})}
  \\ [-.7ex] $#1$}}

\def\mup        {{\ensuremath{\Pmu^+}}\xspace}
\def\mun        {{\ensuremath{\Pmu^-}}\xspace} % muon negative (\mum is taken)

\def\mumu       {{\ensuremath{\Pmu^+\Pmu^-}}\xspace}

\def\g      {{\ensuremath{\Pgamma}}\xspace}

\def\squark    {{\ensuremath{\Ps}}\xspace}

\def\cquark    {{\ensuremath{\Pc}}\xspace}

\def\bquark    {{\ensuremath{\Pb}}\xspace}

\def\pion   {{\ensuremath{\Ppi}}\xspace}
\def\piz    {{\ensuremath{\pion^0}}\xspace}
\def\pip    {{\ensuremath{\pion^+}}\xspace}
\def\pim    {{\ensuremath{\pion^-}}\xspace}

\def\kaon    {{\ensuremath{\PK}}\xspace}
\def\KorKbar {\kern \thebaroffset\optbar{\kern -\thebaroffset \PK}{}\xspace}

\def\Kp      {{\ensuremath{\kaon^+}}\xspace}
\def\Km      {{\ensuremath{\kaon^-}}\xspace}

\def\KS      {{\ensuremath{\kaon^0_{\mathrm{S}}}}\xspace}

\def\Kstarp  {{\ensuremath{\kaon^{*+}}}\xspace}

\def\D       {{\ensuremath{\PD}}\xspace}

\def\DorDbar {\kern \thebaroffset\optbar{\kern -\thebaroffset \PD}\xspace}
\def\Dz      {{\ensuremath{\D^0}}\xspace}

\def\Dp      {{\ensuremath{\D^+}}\xspace}
\def\Dm      {{\ensuremath{\D^-}}\xspace}

\def\DpDm    {\ensuremath{\Dp {\kern -0.16em \Dm}}\xspace}

\def\Dstarp  {{\ensuremath{\D^{*+}}}\xspace}

\def\Ds      {{\ensuremath{\D^+_\squark}}\xspace}

\def\B       {{\ensuremath{\PB}}\xspace}

\def\BorBbar {\kern \thebaroffset\optbar{\kern -\thebaroffset \PB}\xspace}

\def\Bd      {{\ensuremath{\B^0}}\xspace}

\def\BdorBdbar {\kern \thebaroffset\optbar{\kern -\thebaroffset \Bd}\xspace}
\def\Bu      {{\ensuremath{\B^+}}\xspace}

\def\Bs      {{\ensuremath{\B^0_\squark}}\xspace}

\def\BsorBsbar {\kern \thebaroffset\optbar{\kern -\thebaroffset \Bs}\xspace}

\def\jpsi     {{\ensuremath{{\PJ\mskip -3mu/\mskip -2mu\Ppsi}}}\xspace}
\def\psitwos  {{\ensuremath{\Ppsi{(2\PS)}}}\xspace}

\def\chicone  {{\ensuremath{\Pchi_{\cquark 1}}}\xspace}
\def\chictwo  {{\ensuremath{\Pchi_{\cquark 2}}}\xspace}

\def\Y#1S{\ensuremath{\PUpsilon{(#1S)}}\xspace}

\def\proton      {{\ensuremath{\Pp}}\xspace}

\def\Lz          {{\ensuremath{\PLambda}}\xspace}

\def\LorLbar     {\kern \thebaroffset\optbar{\kern -\thebaroffset \PLambda}\xspace}

\def\Xires       {{\ensuremath{\PXi}}\xspace}

\def\Lc          {{\ensuremath{\Lz^+_\cquark}}\xspace}

\def\Lb           {{\ensuremath{\Lz^0_\bquark}}\xspace}

\def\Xibm         {{\ensuremath{\Xires^-_\bquark}}\xspace}

\def\BF         {{\ensuremath{\mathcal{B}}}\xspace}
\def\BR         {\BF}

\newcommand{\decay}[2]{\ensuremath{#1\!\to #2}\xspace} 

\def\to                 {\ensuremath{\rightarrow}\xspace}

\def\AT#1     {\ensuremath{A_{\mathrm{T}}^{#1}}\xspace}           % 2

\def\C#1      {\ensuremath{\mathcal{C}_{#1}}\xspace}                       % 9
\def\Cp#1     {\ensuremath{\mathcal{C}_{#1}^{'}}\xspace}                    % 7
\def\Ceff#1   {\ensuremath{\mathcal{C}_{#1}^{\mathrm{(eff)}}}\xspace}        % 9  
\def\Cpeff#1  {\ensuremath{\mathcal{C}_{#1}^{'\mathrm{(eff)}}}\xspace}       % 7
\def\Ope#1    {\ensuremath{\mathcal{O}_{#1}}\xspace}                       % 2
\def\Opep#1   {\ensuremath{\mathcal{O}_{#1}^{'}}\xspace}                    % 7

\newcommand{\nospaceunit}[1]{\ensuremath{\text{#1}}}       
\newcommand{\aunit}[1]{\ensuremath{\text{\,#1}}}       
\newcommand{\tev}{\aunit{Te\kern -0.1em V}\xspace}
\newcommand{\gev}{\aunit{Ge\kern -0.1em V}\xspace}
\newcommand{\mev}{\aunit{Me\kern -0.1em V}\xspace}
\newcommand{\kev}{\aunit{ke\kern -0.1em V}\xspace}
\newcommand{\ev}{\aunit{e\kern -0.1em V}\xspace}
 
\newcommand{\mevc}{\ensuremath{\aunit{Me\kern -0.1em V\!/}c}\xspace}
\newcommand{\gevc}{\ensuremath{\aunit{Ge\kern -0.1em V\!/}c}\xspace}
\newcommand{\mevcc}{\ensuremath{\aunit{Me\kern -0.1em V\!/}c^2}\xspace}
\newcommand{\gevcc}{\ensuremath{\aunit{Ge\kern -0.1em V\!/}c^2}\xspace}
\def\mm   {\aunit{mm}\xspace}

\def\mum  {\ensuremath{\,\upmu\nospaceunit{m}}\xspace}

\def\fb   {\ensuremath{\aunit{fb}}\xspace}
\def\invfb   {\ensuremath{\fb^{-1}}\xspace}

\newcommand{\chisq}{\ensuremath{\chi^2}\xspace}

\newcommand{\chisqip}{\ensuremath{\chi^2_{\text{IP}}}\xspace}

\def\gsim{{~\raise.15em\hbox{$>$}\kern-.85em
          \lower.35em\hbox{$\sim$}~}\xspace}
\def\lsim{{~\raise.15em\hbox{$<$}\kern-.85em
          \lower.35em\hbox{$\sim$}~}\xspace}

\def\sPlot{\mbox{\em sPlot}\xspace}

\def\pt         {\ensuremath{p_{\mathrm{T}}}\xspace}

\def\ptot       {\ensuremath{p}\xspace}

\def\evtgen     {\mbox{\textsc{EvtGen}}\xspace}

\def\geant      {\mbox{\textsc{Geant4}}\xspace}

\def\photos     {\mbox{\textsc{Photos}}\xspace}

\def\pythia     {\mbox{\textsc{Pythia}}\xspace}

\def\tell1  {TELL1\xspace}
\def\ukl1   {UKL1\xspace}

%% file: local_symbols.tex
\usepackage{rotating}
\usepackage{dashrule}
\usetikzlibrary{patterns}

\def\RPiK{{\ensuremath{\mathcal{R}_{\pion/\kaon}}}\xspace}
\def\RTwoOnePi{{\ensuremath{\mathcal{R}_{2/1}^{\pion}}}\xspace}
\def\RTwoOneK{{\ensuremath{\mathcal{R}_{2/1}^{\kaon}}}\xspace}
\def\thPKm{{\ensuremath{\Ptheta_{\PKm}}}\xspace}
\def\costhPKm{{\ensuremath{\cos\thPKm}}\xspace}

\def\mmc{{\ensuremath{\mm/c}}\xspace}

\def\p{{\ensuremath{\proton}}\xspace}
\def\g{{\ensuremath{\Pgamma}}\xspace}
\def\chicj{{\ensuremath{\Pchi_{\cquark \PJ}}}\xspace}
\def\JpsiG{{\ensuremath{\jpsi\g}}\xspace}
\def\pp{{\ensuremath{\p\p}}\xspace}
\def\PPim{{\ensuremath{\p\pim}}\xspace}
\def\PKm{{\ensuremath{\p\Km}}\xspace}
\def\JpsiPPi{{\ensuremath{\jpsi\p\pim}}\xspace}
\def\JpsiPK{{\ensuremath{\jpsi\p\Km}}\xspace}
\def\ChiconePPi{{\ensuremath{\chicone\p\pim}}\xspace}
\def\ChiconePK{{\ensuremath{\chicone\p\Km}}\xspace}
\def\ChiconeP{{\ensuremath{\chicone\p}}\xspace}
\def\ChiconePim{{\ensuremath{\chicone\pim}}\xspace}
\def\PPim{{\ensuremath{\p\pim}}\xspace}

\def\PizGG{\decay{\piz}{\g\g}}
\def\JpsiMuMu{\decay{\jpsi}{\mumu}}
\def\ChiconeJpsiG{\decay{\chicone}{\jpsi\g}}
\def\ChictwoJpsiG{\decay{\chictwo}{\jpsi\g}}
\def\ChicjJpsiG{\decay{\chicj}{\jpsi\g}}
\def\ChicjJpsiG{\decay{\chicj}{\jpsi\g}}
\def\LzPPim{\decay{\Lz}{\p\pim}}

\def\LbChiconePPi{\decay{\Lb}{\chicone\proton\pim}}
\def\LbChictwoPPi{\decay{\Lb}{\chictwo\proton\pim}}
\def\LbChicjPPi{\decay{\Lb}{\chicj\proton\pim}}
\def\LbChiconePK{\decay{\Lb}{\chicone\proton\Km}}
\def\LbChictwoPK{\decay{\Lb}{\chictwo\proton\Km}}
\def\LbChicjPK{\decay{\Lb}{\chicj\proton\Km}}
\def\LbJpsiPPi{\decay{\Lb}{\jpsi\proton\pim}}
\def\LbJpsiPK{\decay{\Lb}{\jpsi\proton\Km}}

\def\BdJpsiKpPim{\decay{\Bd}{\jpsi\Kp\pim}}
\def\BsJpsiKpKm{\decay{\Bs}{\jpsi\Kp\Km}}

\def\BuJpsiKp{\decay{\Bu}{\jpsi\Kp}}
\def\BuPsiKp{\decay{\Bu}{\psitwos\Kp}}
\def\yieldLbChiconePPi{\ensuremath{105\pm16}\xspace}
\def\yieldLbChictwoPPi{\ensuremath{51\pm16}\xspace}
\def\yieldLbChiconePK{\ensuremath{3133 \pm75}\xspace}
\def\yieldLbChictwoPK{\ensuremath{1766\pm71}\xspace}
\def\signifLbChiconePPi{\ensuremath{9.6}\xspace}
\def\signifLbChictwoPPi{\ensuremath{3.8}\xspace}
\def\effPiK{\ensuremath{1.969\pm0.008}\xspace}
\def\effTwoOnePi{\ensuremath{1.088\pm0.007}\xspace}
\def\effTwoOneK{\ensuremath{1.043\pm0.007}\xspace}

\def\valueRTwoOnePi{{\ensuremath{0.95 \pm 0.30 \pm 0.04 \pm 0.04}}\xspace}
\def\valueRTwoOneK{{\ensuremath{1.06 \pm 0.05 \pm 0.04 \pm 0.04}}\xspace}
\def\valueRTwoOneKPrevious{{\ensuremath{1.02 \pm 0.10 \pm 0.02 \pm 0.05}}\xspace}

\def\systFitmodelRPiK{{\ensuremath{2.4}}\xspace}
\def\systFitmodelRTwoOnePi{{\ensuremath{3.7}}\xspace}
\def\systFitmodelRTwoOneK{{\ensuremath{3.7}}\xspace}

\def\systTrigger{{\ensuremath{1.1}}\xspace}
\def\systDaSiAgr{{\ensuremath{2.0}}\xspace}
\def\systSSSPiK{{\ensuremath{0.4}}\xspace}
\def\systSSSTwoOnePi{{\ensuremath{0.6}}\xspace}
\def\systSSSTwoOneK{{\ensuremath{0.7}}\xspace}
\def\systPiK{{\ensuremath{3.3}}\xspace}
\def\systTwoOnePi{{\ensuremath{3.8}}\xspace}
\def\systTwoOneK{{\ensuremath{3.8}}\xspace}

%% file: title-LHCb-PAPER.tex
% ===============================================================================
% Purpose: LHCb-PAPER journal paper title page template
% Author: 
% Created on: 2010-09-25
% ===============================================================================

%%%%%%%%%%%%%%%%%%%%%%%%%
%%%%%  TITLE PAGE  %%%%%%
%%%%%%%%%%%%%%%%%%%%%%%%%
\begin{titlepage}
\pagenumbering{roman}

% Header ---------------------------------------------------
\vspace*{-1.5cm}
\centerline{\large EUROPEAN ORGANIZATION FOR NUCLEAR RESEARCH (CERN)}
\vspace*{1.5cm}
\noindent
\begin{tabular*}{\linewidth}{lc@{\extracolsep{\fill}}r@{\extracolsep{0pt}}}
\ifthenelse{\boolean{pdflatex}}% Logo format choice
{\vspace*{-1.5cm}\mbox{\!\!\!\includegraphics[width=.14\textwidth]{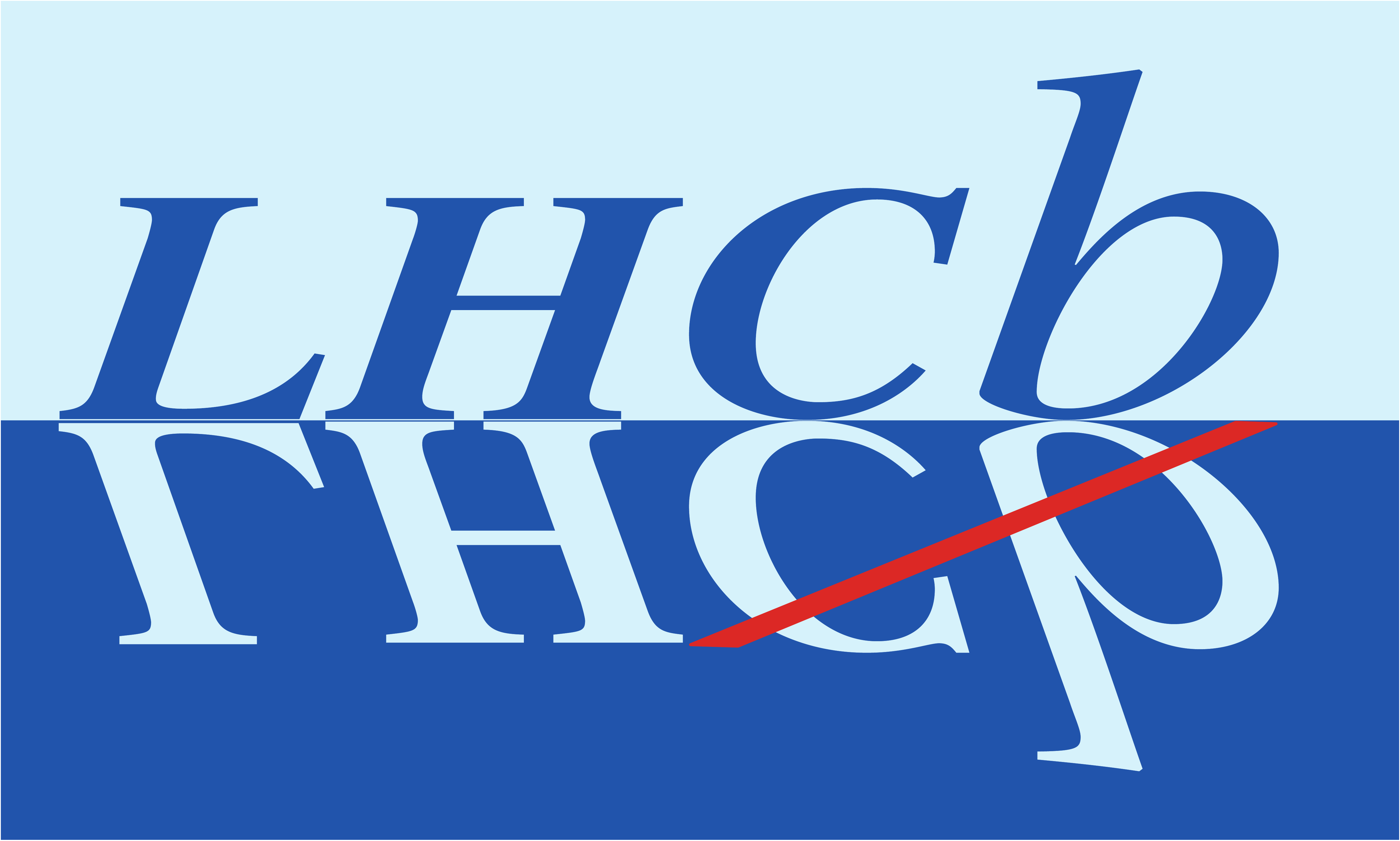}} & &}%
{\vspace*{-1.2cm}\mbox{\!\!\!\includegraphics[width=.12\textwidth]{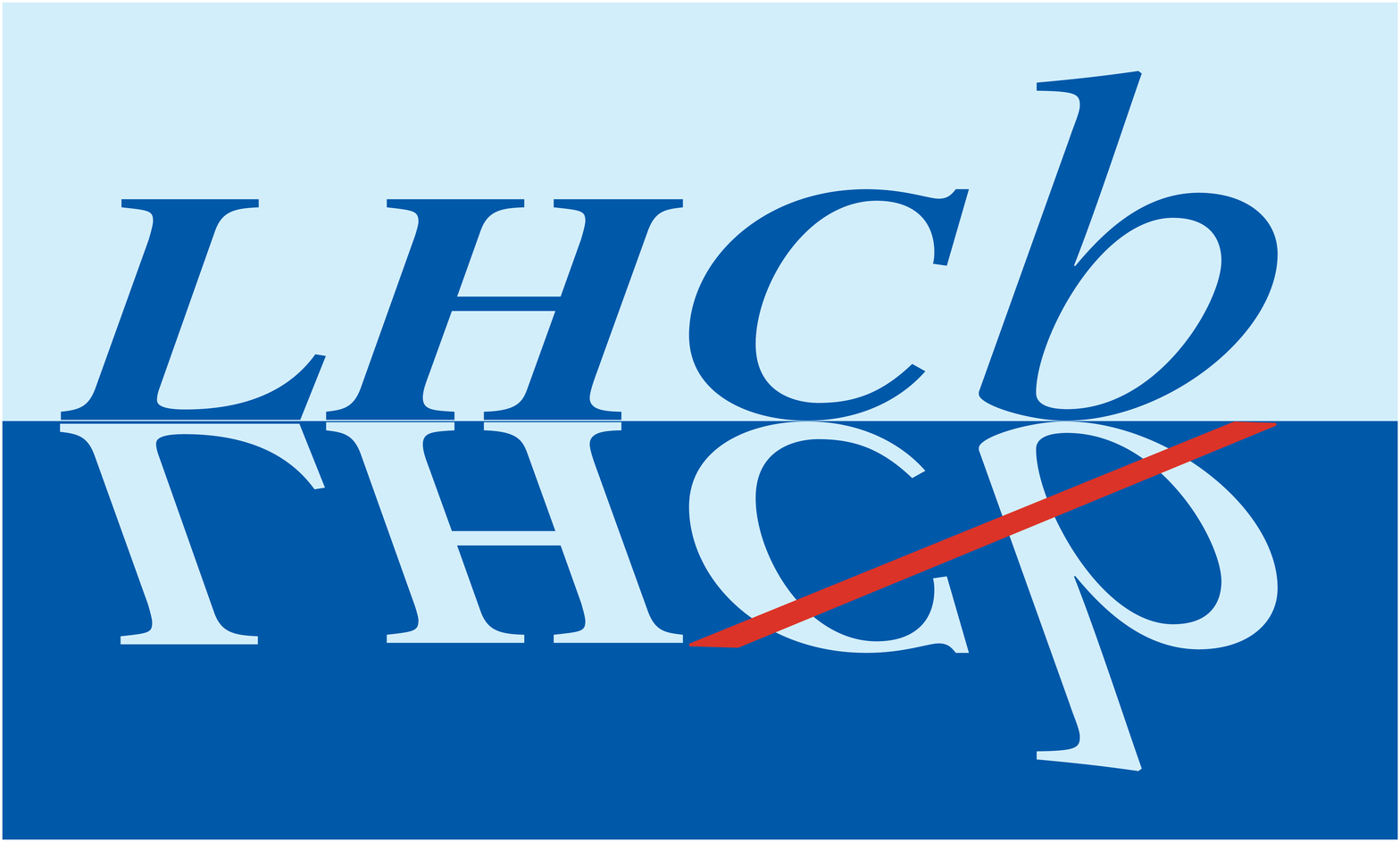}} & &}%
\\
 & & CERN-EP-2021-037 \\  % ID 
 & & LHCb-PAPER-2021-003 \\  % ID 
 & & March 8, 2021 
 %% & & \today \\ % Date - Can also hardwire e.g.: 23 March 2010
 %% & & v1.6 \\
% not in paper \hline
\end{tabular*}

\vspace*{2.0cm}

% Title --------------------------------------------------
{\normalfont\bfseries\boldmath\huge
\begin{center}
% DO NOT EDIT HERE. Instead edit macro in main.tex to keep metadata correct
  \papertitle 
\end{center}
}

\vspace*{1.5cm}

% Authors -------------------------------------------------
\begin{center}
%In the footnote, replace 'paper' by 'Letter' in case of submission to PRL or PLB 
% Edit macro in main.tex to keep metadata correct
\paperauthors\footnote{Authors are listed at the end of this paper.}
\end{center}

\vspace{\fill}

% Abstract -----------------------------------------------
\begin{abstract}
	\noindent
	The Cabibbo\nobreakdash-suppressed decay
	\LbChiconePPi
	%% ~and an~evidence for the~\LbChictwoPPi~decays are done 
	is observed  for the first time
	using data from proton-proton collisions 
	corresponding 
	to an integrated luminosity of~$6\invfb$, 
	collected with the~LHCb detector at a centre-of-mass energy of $13\tev$.
	Evidence for 
	the~\mbox{$\LbChictwoPPi$}~decay is also found. 
	Using the~\mbox{\LbChiconePK}~decay %% with similar topology 
   as normalisation channel, 
    the~ratios  of branching fractions are measured to be
\begin{eqnarray*}
	\dfrac{\BR\left(\LbChiconePPi\right)}{\BR\left(\LbChiconePK\right)}
	& = & \left( 6.59 \pm 1.01 \pm 0.22 \right)\times 10^{-2} \,, %%  \valueRPiK\,,
\\
\dfrac{\BR\left(\LbChictwoPPi\right)}{\BR\left(\LbChiconePPi\right)}
& = & \,\,\valueRTwoOnePi\,,
\\
\dfrac{\BR\left(\LbChictwoPK\right)}{\BR\left(\LbChiconePK\right)}
& = & \,\,\valueRTwoOneK\,,
\end{eqnarray*}
where the~first uncertainty is statistical, 
the~second is systematic and the~third 
is due to the~uncertainties in the~branching fractions 
of~\mbox{$\decay{\Pchi_{\cquark1,2}}{\jpsi\g}$}~decays.

\end{abstract}

\vspace*{2.0cm}

\begin{center}
  For submission to  JHEP
\end{center}

\vspace{\fill}

{\footnotesize 
% Edit macro in main.tex to keep metadata correct
\centerline{\copyright~\papercopyright. \href{\paperlicenceurl}{\paperlicence}.}}
\vspace*{2mm}

\end{titlepage}

%%%%%%%%%%%%%%%%%%%%%%%%%%%%%%%%
%%%%%  EOD OF TITLE PAGE  %%%%%%
%%%%%%%%%%%%%%%%%%%%%%%%%%%%%%%%

%  empty page follows the title page ----
\newpage
\setcounter{page}{2}
\mbox{~}
%\newpage
%
%% Author List ----------------------------
%%  You need to get a new author list!
%\input{LHCb_authorlist.tex}
%
%The author list for journal publications is provided by the Membership Committee shortly after 'approval to go to paper' has been given.
%%It will be made available on the page
%%\verb!http://www.physik.uzh.ch/~strauman/forMemCo/LHCb-PAPER-XXXX-XXX/! .
%It will be sent to you by email shortly after a paper number has beens assigned.
%The author list should be included already at first circulation, 
%to allow new members of the collaboration to verify whether they have been included correctly.
%Occasionally a misspelled name is corrected or associated institutions become full members.
%In that case, a new author list will be sent to you.
%In case line numbering doesn't work well after including the authorlist, try moving the \verb!\bigskip! after the last author to a separate line.
%
%
%The authorship for Conference Reports should be ``The LHCb
%  collaboration'', with a footnote giving the name(s) of the contact
%  author(s), but without the full list of collaboration names.

%% file: introduction.tex
\section{Introduction}
\label{sec:Introduction}

The~amplitude analyses of the~beauty\nobreakdash-baryon decays~\mbox{$\decay{\Lb}{\jpsi\proton\Km}$}
established the~existence of a~new class of
baryonic resonances in the~$\jpsi\proton$~system, 
hidden-charm pentaquarks, 
that cannot be described  within 
the~simplest  pattern of baryon structure  
consisting of three constituent 
quarks~\cite{LHCb-PAPER-2015-029,LHCb-PAPER-2016-009,
LHCb-PAPER-2019-014}.
Evidence for a~pentaquark 
contribution in the~same $\jpsi\proton$~mass region was obtained 
in the~study of 
the~Cabibbo\nobreakdash-suppressed decays 
\mbox{$\decay{\Lb}{\jpsi\proton\pim}$}~\cite{LHCb-PAPER-2016-015}. 
Recently, further evidence
for a~new pentaquark candidate 
in the~\mbox{$\decay{\Xibm}{\jpsi\Lz\Km}$}~decay
has been  reported~\cite{LHCb-PAPER-2020-039}.
Up to now, such hidden\nobreakdash-charm pentaquark
resonances have been observed only in 
the~$\jpsi\proton$ and $\jpsi\Lz$~systems. 
Investigation of such resonances in 
other decay modes, such as
$\Peta_{\cquark}\proton$, 
$\chicone\proton$ and 
$\chictwo\proton$  could shed 
light on the nature of these exotic states. 

The~partial widths of 
the~\mbox{$\decay{\Lb}{\chicone\proton\Km}$}
and~\mbox{$\decay{\Lb}{\chictwo\proton\Km}$}~decays
are measured to be almost equal~\cite{LHCb-PAPER-2017-011}. 
For beauty mesons a different pattern is observed. 
The~known partial  widths for
the~\mbox{$\decay{\B}{\chicone\kaon^{(\ast)}}$} and  
the~\mbox{$\decay{\B}{\chictwo\kaon^{(\ast)}}$}~decays~\cite{
Mizuk:2008me,Aubert:2008ae,LHCb-PAPER-2013-024} 
exhibit
a~large suppression of the~decay modes with the $\chictwo$~state 
with respect to the~$\chicone$~state.
Such suppression agrees with 
expectations from QCD 
factorisation~\cite{Beneke:2008pi}.
More information on the~decays of beauty baryons
to the~\chicone and \chictwo~states
is needed to clarify
the~role of QCD 
factorisation in baryon decays.

In this paper, a~search for the~\LbChiconePPi~and 
\LbChictwoPPi~decays is 
reported, where the~$\chicone$ and $\chictwo$~mesons
are reconstructed via their radiative decays $\decay{ \Pchi_{\cquark1,2}}{\jpsi\g}$,
and the~$\jpsi$~mesons are reconstructed in the~\mumu final state.
The~\mbox{\LbChiconePK}~decay mode, which has a~similar topology, 
is used as normalisation channel.
The~study is based on proton\nobreakdash-proton~(\pp) collision data, 
corresponding to an~integrated luminosity of $6\invfb$, 
collected with the~LHCb detector 
at a~centre\nobreakdash-of\nobreakdash-mass energy of~$13\tev$.
Throughout this~paper the inclusion 
of charge\nobreakdash-conjugated processes 
is implied and the~symbol \chicj~is used to 
denote the $\chicone$~and \chictwo~states collectively.

%% file: detector.tex
\section{Detector and simulation}
\label{sec:Detector}

The \lhcb detector~\cite{LHCb-DP-2008-001,LHCb-DP-2014-002} is 
a~single\nobreakdash-arm forward spectrometer 
covering the~pseudorapidity range \mbox{$2<\eta <5$}, 
designed for the study of particles containing \bquark~or \cquark~quarks.
The~detector includes a~high\nobreakdash-precision tracking system consisting
of a~silicon\nobreakdash-strip vertex detector surrounding 
the~$\proton\proton$~interaction region, 
a~large\nobreakdash-area silicon\nobreakdash-strip detector 
located upstream of a~dipole magnet with a~bending power of 
about~$4{\mathrm{\,Tm}}$, and three stations of 
silicon\nobreakdash-strip detectors 
and straw drift tubes placed downstream of the~magnet.
The~tracking system provides a~measurement of the~momentum, \ptot, 
of charged particles with a~relative uncertainty that varies from 
0.5\% at low momentum to 1.0\% at~200\gevc.
The~minimum distance of a~track to a~primary 
$\proton\proton$ collision vertex\,(PV), 
the~impact parameter\,(IP), 
is measured with a~resolution of~$(15+29/\pt)\mum$, 
where \pt~is the~component of the~momentum transverse 
to the~beam, in\,\gevc.
Different types of charged hadrons are distinguished using information
from two ring\nobreakdash-imaging Cherenkov\,(RICH) detectors.
Photons, electrons and hadrons are identified by 
a~calorimeter system consisting of scintillating\nobreakdash-pad 
and preshower detectors, an~electromagnetic
% calorimeter
and a~hadronic calorimeter~\cite{LHCb-DP-2020-001}.
Muons are identified by a~system composed of alternating layers of iron and 
multiwire proportional chambers.

The~online event selection is performed by a~trigger, 
which consists of a~hardware stage, based on information 
from the~calorimeter and muon systems, 
followed by a~software stage, which applies a~full event reconstruction.
At~the~hardware trigger stage, events are required to have 
a~muon with high transverse momentum or dimuon candidates 
in which a~product of the~\pt of the~muons 
has a~high value. 
%% with 
%% a~high value of the~product of the~\pt of the~muons.
In~the~software trigger, two oppositely charged muons are required 
to form a~good\nobreakdash-quality vertex that is significantly 
displaced from every~PV, with a~dimuon mass exceeding $2.7\gevcc$.

Simulated events are used to describe signal shapes 
and to compute the~efficiencies needed to determine 
the~branching fraction ratios.
In~the~simulation, $\proton\proton$~collisions are 
generated using \pythia~\cite{Sjostrand:2007gs} with 
a~specific \lhcb configuration~\cite{LHCb-PROC-2010-056}.
Decays of unstable particles are 
described by \evtgen~\cite{Lange:2001uf}, 
in which final\nobreakdash-state radiation is generated 
using \photos~\cite{%%Golonka:2005pn,
davidson2015photos}.
The~interaction of the~generated particles with the~detector, 
and its response, are implemented using 
the~\geant toolkit~\cite{Allison:2006ve, *Agostinelli:2002hh} 
as described in Ref.~\cite{LHCb-PROC-2011-006}.
The~transverse momentum  and rapidity spectra of 
the~\Lb~baryons in simulated samples are adjusted 
to match those observed 
in a~high\nobreakdash-yield low\nobreakdash-background 
sample of reconstructed \mbox{\LbJpsiPK}~decays.
In~the~simulation, the~\Lb~baryon decays are produced according
to a~phase space decay model.
Simulated \LbChicjPK decays are corrected to reproduce 
the~\PKm~mass and \costhPKm~distributions observed in data, 
where \thPKm is the~helicity angle of the~\PKm~system, 
defined as the~angle between the~momentum vectors 
of the~kaon and the \Lb~baryon in the~\PKm rest frame.
Large calibration samples of low-background 
decays
\mbox{$\decay{\Dstarp}{ \left( \decay{\Dz}{\Km\pip}\right) \pip}$},
\mbox{$\decay{\KS}{\pip\pim}$},
\mbox{$\decay{\Ds}{ \left( \decay{\Pphi}{\Kp\Km}\right) \pip}$},
\mbox{$\decay{\Lz}{\proton\pim}$} and 
\mbox{$\decay{\Lc}{\proton\Km\pip}$}~\cite{LHCb-DP-2012-003,LHCb-DP-2018-001}
are used to resample the~combined detector response used 
for the~identification of protons, kaons and pions.
To~account for imperfections in the~simulation of 
charged\nobreakdash-particle reconstruction, 
the~track reconstruction efficiency determined from simulation 
is corrected using control channels in data~\cite{LHCb-DP-2013-002}.

%% file: selection.tex
\section{Event selection}

The~signal \mbox{\LbChicjPPi} and the~normalisation 
\mbox{\LbChicjPK} decays are both reconstructed 
using the~decay modes \mbox{\ChicjJpsiG} and \mbox{\JpsiMuMu}.
A~loose preselection similar to that used 
in Refs.~\cite{LHCb-PAPER-2013-024,
LHCb-PAPER-2014-008,
LHCb-PAPER-2015-060,
LHCb-PAPER-2018-022,
LHCb-PAPER-2019-023} is applied, 
followed by a~multivariate classifier based on a~decision tree with gradient boosting\,(BDTG)~\cite{BDTG}.

Muon, proton, pion and kaon candidates are 
identified combining information 
from the~RICH, calorimeter and muon detectors.
They are required to have 
transverse momenta larger 
than 550, 500, 200~and 200\mevc, respectively.
To~ensure efficient particle identification, 
kaons and pions are required to have a~momentum 
between \mbox{3.2~and 150\gevc}, whilst protons must have 
momentum between 10~and 150\gevc.
To~reduce the combinatorial background due to 
particles produced in \pp~interactions, 
only tracks that are inconsistent 
with originating from any~PV are used.

Pairs of oppositely charged muons consistent with originating 
from a~common vertex are combined to 
form~\mbox{\JpsiMuMu} candidates.
The~transverse momentum of the~dimuon candidate 
is required to be in excess of 2\gevc, and 
the~mass of the~$\mup\mun$~system is required to be between 
3.020~and 3.135\gevcc, where the~asymmetric mass range around 
the~known \jpsi~mass~\cite{PDG2020} is chosen to account for 
final\nobreakdash-state radiation.
The~position of the~reconstructed dimuon vertex 
is required to be inconsistent 
with that of any reconstructed PV.

To~create \chicj~candidates,  
the~selected \jpsi~candidates are 
combined with photon candidates that have been 
reconstructed using clusters in the~electromagnetic 
calorimeter.
Only clusters that are not matched 
to the~trajectory of a~track extrapolated 
from the~tracking system to the~cluster 
position in the~electromagnetic calorimeter 
are used in the~analysis~\cite{LHCb-DP-2020-001}.
The~transverse energies of 
the~photon candidates are  required to exceed 
400\mev.
To~suppress the~large combinatorial
background from \PizGG~decays, 
photons that can form a~\mbox{\PizGG}~candidate with 
mass within 25\mevcc of the~known \piz~mass~\cite{PDG2020} 
are ignored~\cite{LHCb-PAPER-2012-022,
LHCb-PAPER-2014-056}.
The~\chicj~candidates are 
selected in the~\JpsiG mass region between \mbox{$3.4$ and~$3.7\gevcc$}.

The~selected \chicj~candidates are combined with \PPim~or 
\PKm~pairs to create 
\mbox{$\decay{\Lb}{\chicj\proton\pim}$}
or 
\mbox{$\decay{\Lb}{\chicj\proton\Km}$}~candidates, respectively.
A~kinematic fit~\cite{Hulsbergen:2005pu} that constrains 
the~four charged final-state particles to form a~common vertex, 
the~mass of the~\mumu~combination 
to equal the~known \jpsi~mass~\cite{PDG2020}
%,the~mass of the~\JpsiG~combination to the~known \chicone~mass
and the~\Lb~candidate to originate from the~associated PV, 
is performed.
Each~\Lb~candidate is associated with 
the~PV that yields the~smallest \chisqip, 
where \chisqip is defined as the~difference 
in the~vertex-fit \chisq of a~given 
PV reconstructed with and without 
the~particle under consideration.
A~good\nobreakdash-quality fit is required 
to further suppress combinatorial background.
In addition, the~measured decay 
time of the~\Lb~candidate, 
calculated with respect to the~associated PV, 
is required to be greater than 0.1\mmc 
to suppress poorly reconstructed candidates 
and background from particles originating directly from the~PV.

To~suppress cross\nobreakdash-feed 
from 
\mbox{$\decay{\Bd}{\chicj\Kp\pim}$}
decays with the~positively charged
kaon\,(negatively charged pion) 
misidentified as a proton\,(antiproton) for 
the~signal\,(normalisation) channel, 
%% a~veto is applied on 
the~\Lb~candidate mass 
recalculated with a~kaon\,(pion) mass hypothesis 
for the~proton is required to be inconsistent with 
the~known $\Bd$~meson mass~\cite{PDG2020}.
In a~similar way, \Lb~candidates are rejected if 
the~mass of the~\PPim\,(\PKm) combination is 
consistent  with the~known $\Pphi$\nobreakdash-meson 
mass~\cite{PDG2020} 
when a~kaon mass hypothesis is used for both hadrons.
To~suppress background from 
the~\LzPPim~decay, candidates with a~\PPim~mass 
that is consistent with the~known 
mass of the~\Lz~baryon~\cite{PDG2020} 
are rejected.
The~contributions from 
the~\LbJpsiPPi and \LbJpsiPK  decays
combined with random photons
are eliminated by 
the~requirement that
the~mass of the~\Lb~candidate 
calculated without a~photon 
is inconsistent with 
the~known
mass of 
the~\Lb~baryon~\cite{LHCb-PAPER-2017-011,PDG2020}.
Finally, 
the~contributions 
from wrongly reconstructed 
%% decays of beauty hadrons 
\mbox{\BdJpsiKpPim}, 
\mbox{\LbJpsiPK} and 
\mbox{\BsJpsiKpKm}~decays, 
combined with random photons,
are rejected by the~requirement 
that the~mass of the~\Lb~candidate 
recalculated 
using different mass hypotheses for 
the~pion, kaon and proton~candidates
and ignoring the~photon 
in the~final state, 
be inconsistent with the~known 
mass of the~corresponding beauty hadron.

To~suppress a~potentially 
large combinatorial background, 
separate BDTG classifiers are used for 
the~\LbChicjPPi~and \LbChicjPK~candidates.
The~classifiers are trained 
using simulated samples of 
\LbChiconePPi and \LbChiconePK decays
as signal. 
The~\mbox{\LbChiconePPi}~and \LbChiconePK~candidates 
with 
the~\ChiconePPi~and \ChiconePK~mass
in the~range 
\mbox{$5.65<  
m_{\ChiconePPi}$} and  
\mbox{$m_{\ChiconePK} 
< 6.00\gevcc$} %%, respectively, 
are used as background.
The~$k$\nobreakdash-fold cross\nobreakdash-validation 
technique~\cite{chopping} 
with $k=7$ is used to avoid 
introducing a~bias in the~BDTG output.
The~BDTG classifier 
for the~\LbChicjPPi\,(\LbChicjPK)~candidates 
is trained 
on variables related to the~reconstruction quality, 
kinematics and  decay time of \Lb~candidates,
kinematics of particles in the~final state
and the~estimated probabilities that protons and pions\,(kaons) 
are correctly identified by the~particle
identification detectors~\cite{%%LHCb-PROC-2011-008,
LHCb-DP-2012-003,LHCb-DP-2018-001}.
The~requirement on the~BDTG output is chosen 
to maximize the~figure\nobreakdash-of\nobreakdash-merit  
${S}/{\sqrt{S+B}}$, where $S$ and $B$ are   
expected signal and background 
yields, correspondingly.
The~signal yields are 
estimated from the~simulated samples, 
normalised  to the~signal yields 
observed in data for the~loose 
requirements on the~BDTG  output, and   
the~background yield $B$
is estimated from the~fit 
to data using a~model, 
described in Sec.~\ref{sec:signals}.

After application of the~BDTG requirement, 
$6\%$ of  events
with $\LbChicjPPi$~candidates 
in  the~\mbox{$5.4<m_{\ChiconePPi}<5.8\gevcc$}~region 
and 13\% of events
with $\LbChicjPK$~candidates 
in the~\mbox{$5.3< m_{\ChiconePK} <5.8\gevcc$}~region
contain multiple candidates.
These multiple candidates are 
predominantly
caused by the~\JpsiPPi or \JpsiPK~combination being 
combined with different photons in the~event.
A~study using simulation 
shows that the~random photons causing 
multiple candidates
typically have lower transverse energy 
with respect to that of 
the~photons originating from the~\Lb~baryon decay.
Therefore, to reduce multiple candidates  
for each event, only 
the~\Lb~candidate with 
the~highest transverse energy photon is retained.

To~improve the~\Lb~mass resolution, 
the~mass of the~\Lb~candidates is calculated using 
a~kinematic fit~\cite{Hulsbergen:2005pu}, 
similar to the~one described above, 
but with an additional constraint 
fixing the~mass of the~\JpsiG~combination 
to the~known \chicone~mass~\cite{PDG2020}.
For~the~\LbChiconePPi and \LbChiconePK~decays, 
the~mass  calculated with such a constraint  
forms a~narrow peak at the~known mass of the~\Lb~baryon, 
while for 
the~\LbChictwoPPi and \LbChictwoPK~decays
the~narrow peak is shifted towards lower values~\cite{LHCb-PAPER-2013-024,LHCb-PAPER-2017-011}. 

%% file: signals.tex
\section{Signal yields and efficiencies}
\label{sec:signals}

\begin{figure}[tb]
\centering
\ifthenelse{\boolean{completeplots}}%% 
{ \includegraphics*[width=0.98\textwidth]{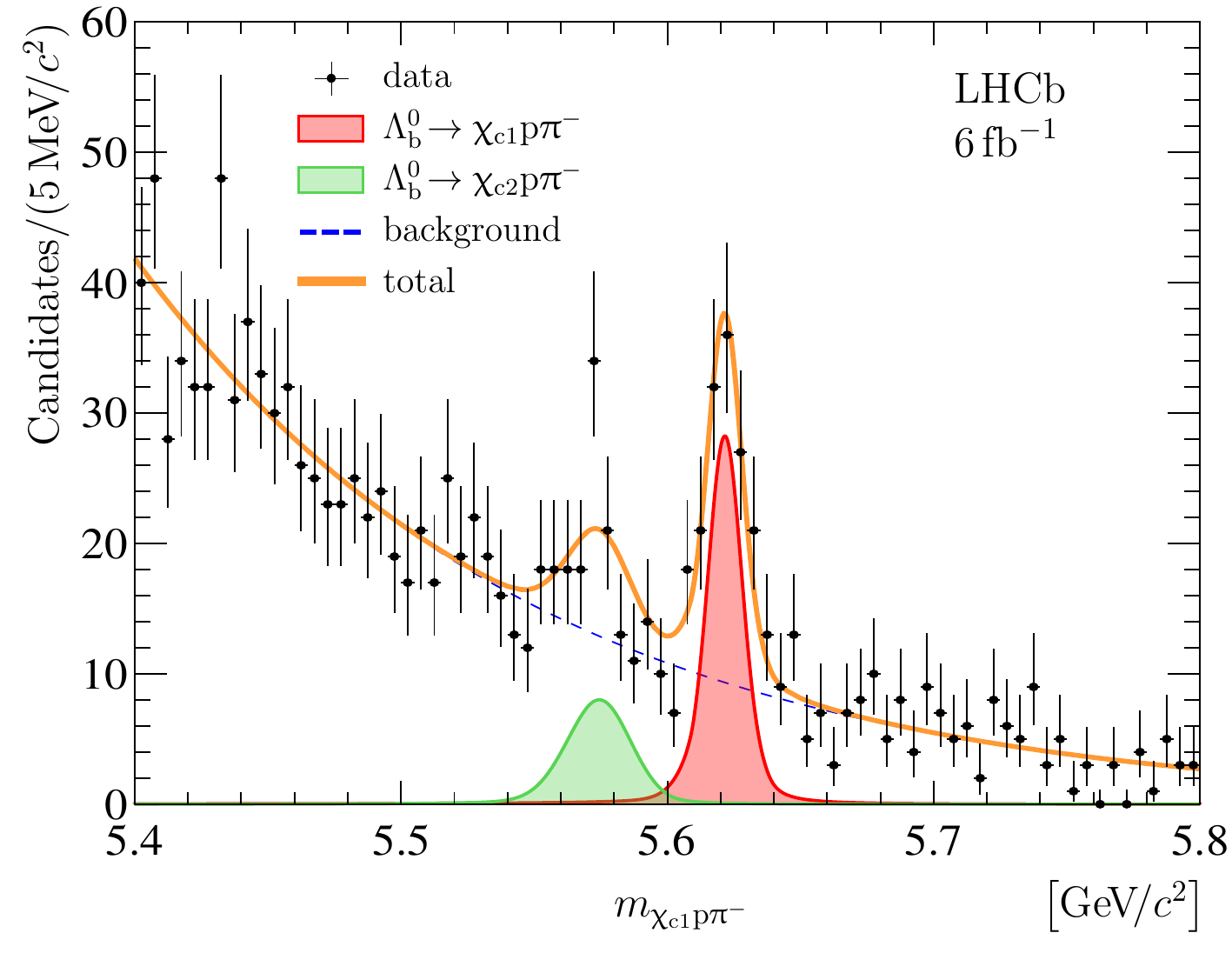} }
{ \setlength{\unitlength}{1mm}
	\begin{picture}(150,111)
	    %% 
        %%\graphpaper[5](-10,-10)(170,130)
        %%
        \definecolor{gr}{rgb}{0.35,0.83,0.33}
        \definecolor{vi}{rgb}{0.8,0.0,1.0}
		\put(0,1){\includegraphics*[width=150mm,height=110mm]{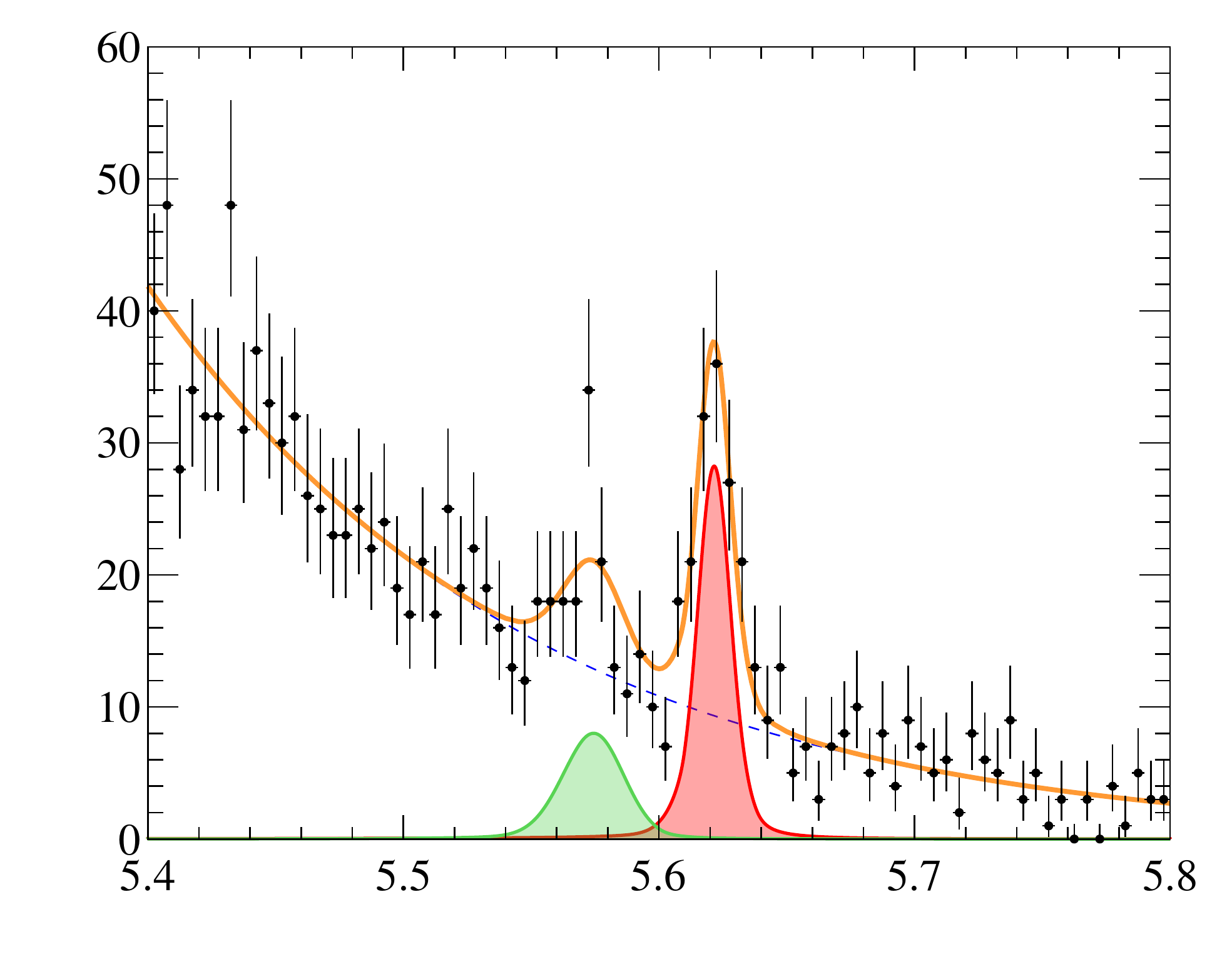}}
		\put(5,56){\large\begin{sideways}Candidates/$(5\mevcc)$\end{sideways}}
		\put(74,1){\large$m_{\ChiconePPi}$}
		\put(124,1){\large$\left[\!\gevcc\right]$}
		\put(112,93){\large$\begin{array}{l}\lhcb \\ 6\invfb\end{array}$}
		\put(39,99){\line(1,0){4}}
        \put(41,97){\line(0,1){4}}
        \put(41,99){\circle*{1.0}}
		\put(37,91.5){\begin{tikzpicture}[x=1mm,y=1mm]\filldraw[fill=red!35!white,draw=red,thick](0,0)rectangle(7.5,3);\end{tikzpicture}}
		\put(37,85.5){\begin{tikzpicture}[x=1mm,y=1mm]\filldraw[fill=gr!35!white,draw=gr,thick](0,0)rectangle(7.5,3);\end{tikzpicture}}
		\put(37,80.8){\color[rgb]{0,0,1}{\hdashrule[0.0ex][x]{8mm}{1.5pt}{2.0mm 0.3mm}}}
		\put(37,74.8){\color[RGB]{255,153,51}{\rule{8mm}{3.0pt}}}
		\put(47,98){data}
		\put(47,92){\LbChiconePPi}
		\put(47,86){\LbChictwoPPi}
		\put(47,80){background}
		\put(47,74){total}
	\end{picture}
	}
	\caption{\small
		Mass distribution for selected~\LbChicjPPi~candidates.
		A~fit, described in the~text, is overlaid. 
	}
	\label{fig:fits_signal}
\end{figure}
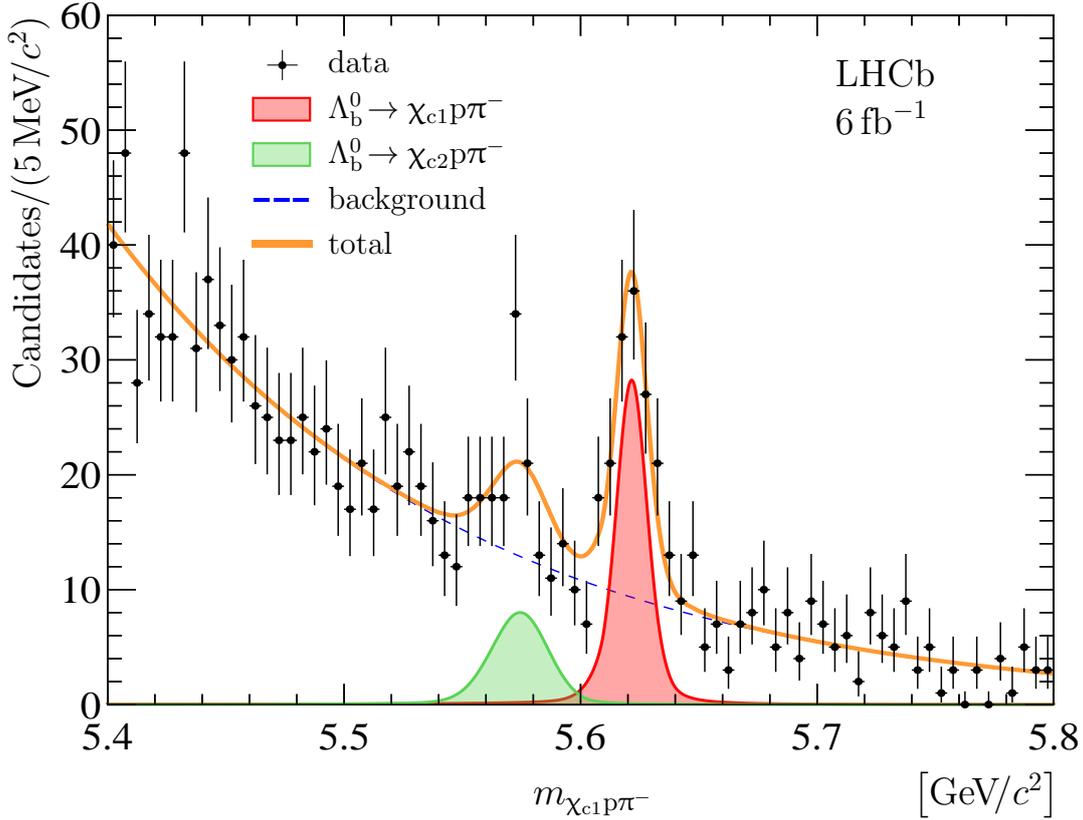

\begin{figure}[tb]
\centering
\ifthenelse{\boolean{completeplots}}%% 
{ \includegraphics*[width=0.98\textwidth]{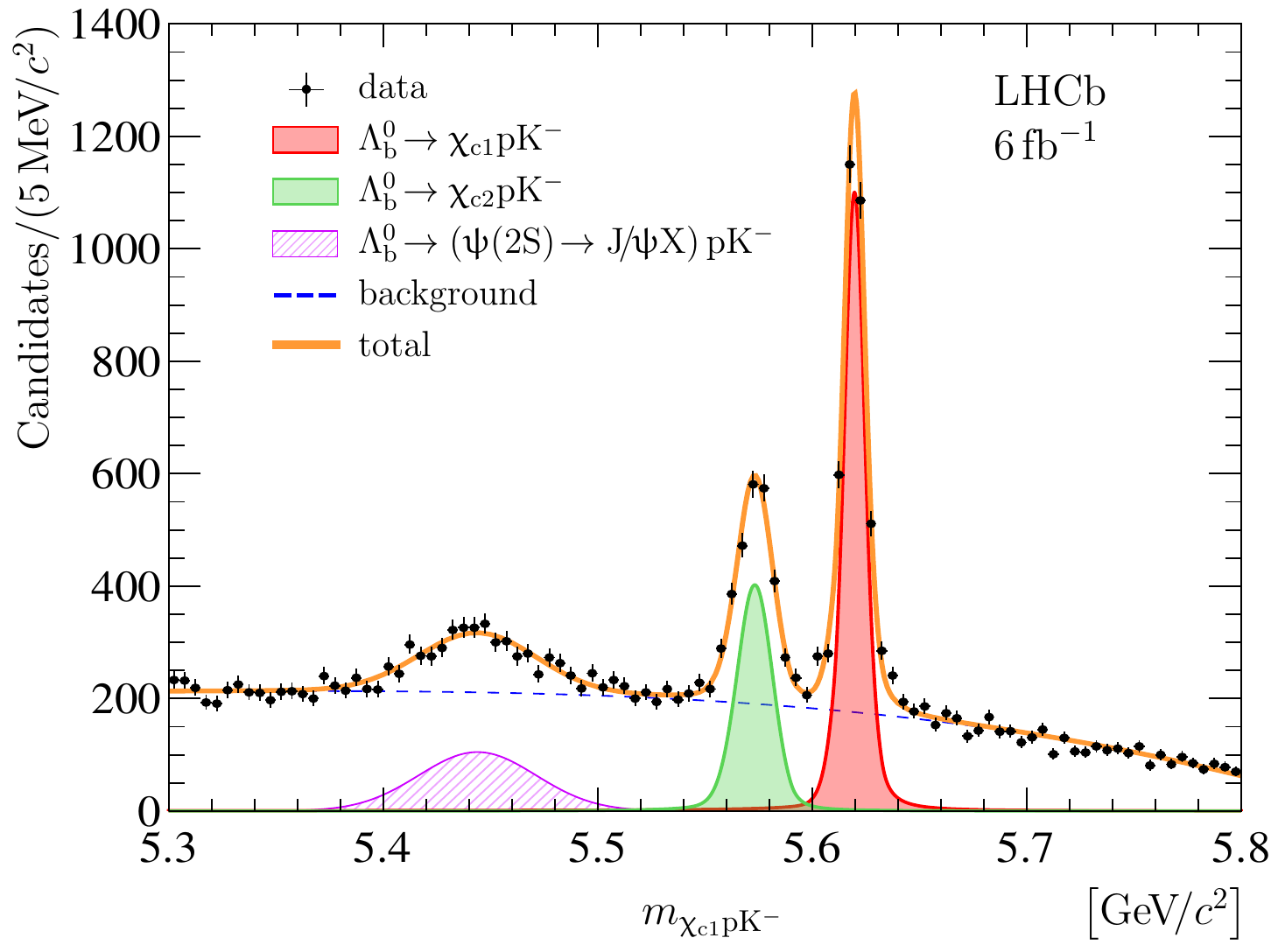} }
{ \setlength{\unitlength}{1mm}
  \begin{picture}(150,111)
	    %%
        %%\graphpaper[5](-10,-10)(170,130)
        %%
        \definecolor{gr}{rgb}{0.35,0.83,0.33}
        \definecolor{vi}{rgb}{0.8,0.0,1.0}
		\put(0,1){\includegraphics*[width=150mm,height=110mm]{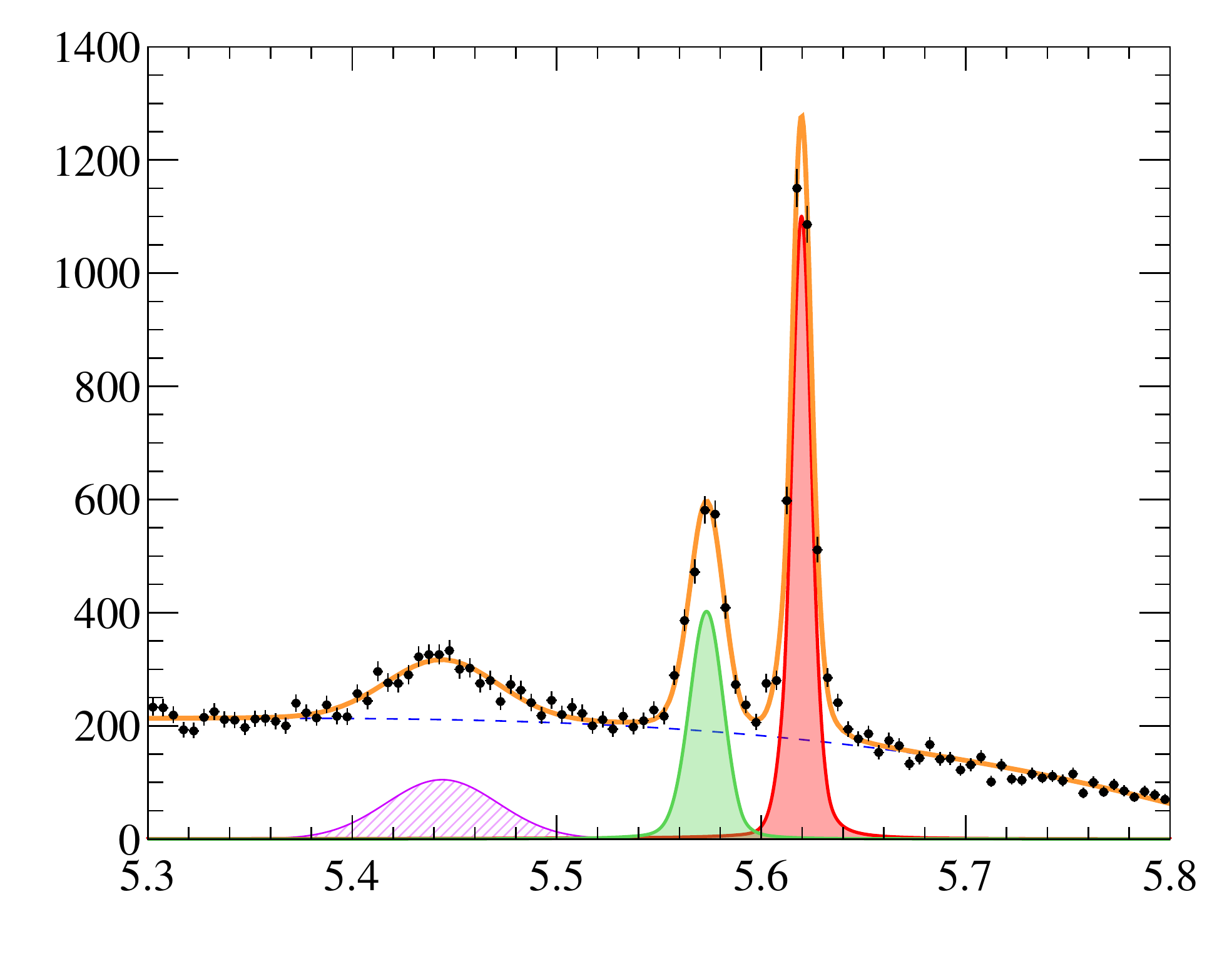}}
		\put(0,56){\large\begin{sideways}Candidates/$(5\mevcc)$\end{sideways}}
		\put(73,1){\large$m_{\ChiconePK}$}
		\put(124,1){\large$\left[\!\gevcc\right]$}
     	\put(112,93){\large$\begin{array}{l}\lhcb \\  6\invfb\end{array}$}
		\put(32,98){\line(1,0){4}}
        \put(34,96){\line(0,1){4}}
        \put(34,98){\circle*{1.0}}
		\put(30,90.5){\begin{tikzpicture}[x=1mm,y=1mm]\filldraw[fill=red!35!white,draw=red,thick](0,0)rectangle(7.5,3);\end{tikzpicture}}
		\put(30,84.5){\begin{tikzpicture}[x=1mm,y=1mm]\filldraw[fill=gr!35!white,draw=gr,thick](0,0)rectangle(7.5,3);\end{tikzpicture}}
	    \put(30,78.5){{\begin{tikzpicture}[x=1mm,y=1mm]\draw[thin, vi, pattern=north east lines, pattern color=vi!35!white] (0,0) rectangle (7.5,3);\end{tikzpicture}}}
		\put(30,73.8){\color[rgb]{0,0,1}{\hdashrule[0.0ex][x]{8mm}{1.5pt}{2.0mm 0.3mm}}}
		\put(30,67.8){\color[RGB]{255,153,51}{\rule{8mm}{3.0pt}}}
		\put(40,97){data}
		\put(40,91){\LbChiconePK}
		\put(40,85){\LbChictwoPK}
		\put(40,79){$\decay{\Lb}{\left(\decay{\psitwos}{\jpsi\PX}\right)\proton\Km}$}
		\put(40,73){background}
		\put(40,67){total}
	\end{picture}
	}
	\caption{\small
		Mass distribution for 
		selected~\LbChicjPK~candidates.
		A~fit, described in the~text, is overlaid. 
	}
	\label{fig:fits_norm}
\end{figure}

The~mass distributions for selected \LbChicjPPi 
and \LbChicjPK candidates
are shown in Figs.~\ref{fig:fits_signal}
and~\ref{fig:fits_norm}, respectively.
The~signal yields are determined using unbinned 
extended maximum\nobreakdash-likelihood 
fits to these distributions.
For~the~\LbChicjPPi channel, 
the fit model consists of 
two signal components, corresponding to 
the~\LbChiconePPi~and \LbChictwoPPi~decays, 
as described below, 
and a~combinatorial background component 
that is described by the~product of 
an~exponential function 
and a~first\nobreakdash-order polynomial function,
required to be positive in the~relevant mass range.
For the~\LbChicjPK channel, 
the~fit model consists 
of two~signal components,
corresponding to 
the~\LbChiconePK~and \LbChictwoPK~decays,
a~combinatorial background 
component which is described by 
a~concave  third\nobreakdash-order positive polynomial function 
and 
a~component from partially reconstructed 
\Lb~baryon decays, such as 
\mbox{$\decay{\Lb}{\psitwos\proton\Km}$}
with subsequent decays
\mbox{$\decay{\psitwos}{\jpsi\Ppi\Ppi}$},
\mbox{$\decay{\psitwos}{\jpsi\Peta}$} or 
\mbox{$\decay{\psitwos}{\left(\decay{\chicone}{\jpsi\g}\right)\g}$},
which is described by a~Gaussian function.
Each of the~four signal components is described by 
the~sum of two Crystal Ball\,(CB) functions~\cite{Skwarnicki:1986xj} 
with a~common mean and power\nobreakdash-law tails on both sides.
The~tail parameters of the~CB functions, 
the~ratio of the~widths of the~two CB functions, 
and their relative normalisation are fixed 
to the~values obtained from simulation.
%% fits to simulated samples.
%% The~positions  for 
%% the~\LbChictwoPPi and \LbChictwoPK
%% components are biased towards 
%% lower  values 
%% with respect to the~known mass of the~\Lb~baryon,
%% due to the~\chicone~mass constraint used in the~calculation 
%% of the~\Lb~candidate mass.
The~widths and the~difference in the~mean values for 
the~large \LbChiconePK and \LbChictwoPK components 
are allowed to vary in the~fit, while for 
the~small \LbChiconePPi and \LbChictwoPPi components,
the~difference in the~mean values and 
the~ratio of widths are constrained 
to the~values obtained from simulation.

The signal yields for
the~\LbChiconePPi, \LbChictwoPPi, \LbChiconePK 
and \LbChictwoPK decay modes
are summarized in Table~\ref{tab:fits}.
The statistical significance
for the~\mbox{\LbChiconePPi} and 
\mbox{\LbChictwoPPi}~fit components is
estimated using Wilks' theorem~\cite{Wilks:1938dza}.
The~significance for 
the~\mbox{\LbChictwoPPi}~signal is
confirmed by simulating a~large number 
of pseudoexperiments according to 
the~background 
distributions observed in data.
The~statistical significance is found to be 
$\signifLbChiconePPi$ 
and $\signifLbChictwoPPi$~standard deviations for 
the~\LbChiconePPi and \LbChictwoPPi decay modes, respectively.

\begin{table}[tb]
	\centering
	\caption{\small 
	Signal yields, $N$, from the~fits described in the~text. 
	The~uncertainties are statistical only. 
	}
	\label{tab:fits}
	\vspace{2mm}
	\begin{tabular*}{0.38\textwidth}{@{\hspace{3mm}}l@{\extracolsep{\fill}}c@{\hspace{3mm}}}
	  Decay mode  &  $N$
   \\[1mm]
  \hline 
  \\[-2mm]
     \LbChiconePPi  & $\phantom{0}\yieldLbChiconePPi$ \\   
     \LbChictwoPPi  & $\phantom{00}\yieldLbChictwoPPi$ \\
     \LbChiconePK   & $\yieldLbChiconePK$  \\
     \LbChictwoPK   & $\yieldLbChictwoPK$  
	\end{tabular*}
\end{table}

The measured yields for the~\mbox{\LbChiconePPi}, 
\mbox{\LbChictwoPPi}, 
\mbox{\LbChiconePK} 
and \mbox{\LbChictwoPK} decay modes are
used to calculate the~ratios 
of branching fractions 
\begin{subequations} \label{eq:R}
\begingroup
\allowdisplaybreaks
\begin{eqnarray}
	\RPiK
	\equiv  
	\dfrac{\BR\left(\LbChiconePPi\right)}{\BR\left(\LbChiconePK\right)}
	 & = & 
	 \dfrac{N_{\LbChiconePPi}}{N_{\LbChiconePK}}
	 \times
	 \dfrac{\upvarepsilon_{\LbChiconePK}}{\upvarepsilon_{\LbChiconePPi}}\,,
	\\
	\RTwoOnePi 
	 \equiv  
	 \dfrac{\BR\left(\LbChictwoPPi\right)}{\BR\left(\LbChiconePPi\right)}
	 & = &  
	 \dfrac{N_{\LbChictwoPPi}}{N_{\LbChiconePPi}}
	 \times
	 \dfrac{\upvarepsilon_{\LbChiconePPi}}{\upvarepsilon_{\LbChictwoPPi}}
	 \times
	 \dfrac{\BR(\ChiconeJpsiG)}{\BR(\ChictwoJpsiG)}\,,
	\\
	\RTwoOneK 
	 \equiv  
	 \dfrac{\BR\left(\LbChictwoPK\right)}{\BR\left(\LbChiconePK\right)}
	  & = &  
	  \dfrac{N_{\LbChictwoPK}}{N_{\LbChiconePK}}
	  \times
	  \dfrac{\upvarepsilon_{\LbChiconePK}}{\upvarepsilon_{\LbChictwoPK}}
	  \times
	  \dfrac{\BR(\ChiconeJpsiG)}{\BR(\ChictwoJpsiG)}\,,
\end{eqnarray}
\endgroup
\end{subequations}
where $N$ stands for the~measured yield,
$\varepsilon$ denotes the~efficiency 
of the~corresponding decay and
\mbox{$\BR(\decay{\chicj}{\jpsi\g})$}~are
the~branching fractions of the~radiative 
\mbox{$\decay{\chicj}{\jpsi\g}$}~decays, 
taken from Ref.~\cite{PDG2020}.
The~efficiency is defined  as the~product of the~detector 
acceptance,  reconstruction, selection and trigger efficiencies, 
where each subsequent efficiency is defined with 
respect to the~previous one. Each of the~partial efficiencies 
is calculated using the~appropriately corrected simulation samples.
The~efficiencies are determined separately for each data\nobreakdash-taking 
period and are combined according to 
the~corresponding luminosity for each period.
The~ratios of the~total efficiencies are determined to be
\begin{subequations}\label{eq:efficiency}
\begingroup
\allowdisplaybreaks
\begin{eqnarray}
	\dfrac{\upvarepsilon_{\LbChiconePK}}{\upvarepsilon_{\LbChiconePPi}} 
	&=&
	\effPiK\,,
	\label{eq:efficiencyPiK}
	\\
	\dfrac{\upvarepsilon_{\LbChiconePPi}}{\upvarepsilon_{\LbChictwoPPi}} 
	&=&
	\effTwoOnePi\,,
	\label{eq:efficiencyTwoOnePi}
	\\
	\dfrac{\upvarepsilon_{\LbChiconePK}}{\upvarepsilon_{\LbChictwoPK}} 
	&=&
	\effTwoOneK\,,
    \label{eq:efficiencyTwoOneK}
\end{eqnarray}
\endgroup
\end{subequations}
where only the~uncertainty that arises from the~sizes of 
the~simulated samples is given.
The~\LbChiconePPi decay channel has a~much higher combinatorial 
background level, 
therefore, the~BDTG selection 
is less efficient with 
respect to that for the~\LbChiconePK channel, 
which is the~main factor causing 
the~difference in total efficiencies 
for these channels. 
Using these ratios of efficiencies and  
the~measured yields from Table~\ref{tab:fits}, 
the~ratios of the~branching fractions are found to be 
\begin{subequations}
\begingroup
\allowdisplaybreaks
\begin{eqnarray*}
\RPiK      & = & \left(6.59 \pm 1.01\right)\times 10^{-2} \,, \\ 
\RTwoOnePi & = & \phantom{(}0.95 \pm 0.30     \,, \\ 
\RTwoOneK  & = & \phantom{(}1.06 \pm 0.05     \,, 
\end{eqnarray*}
\endgroup
\end{subequations}
where the uncertainties are statistical only.
Systematic uncertainties are discussed in the~next section.

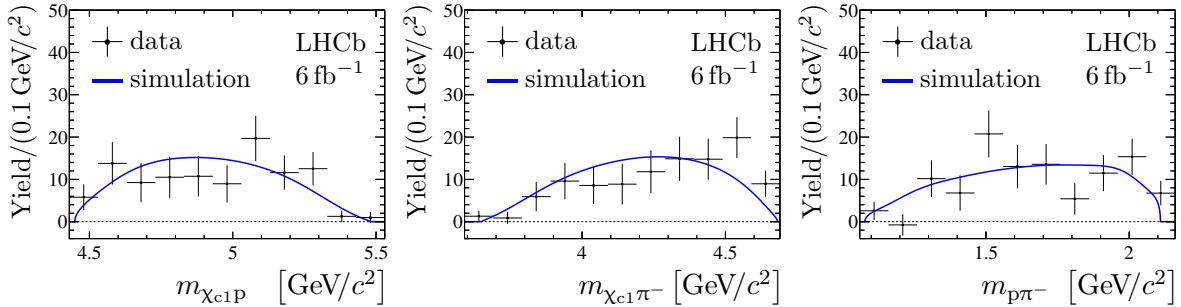
\begin{figure}[tb]
\centering
\ifthenelse{\boolean{completeplots}}%% 
{ \includegraphics*[width=0.98\textwidth]{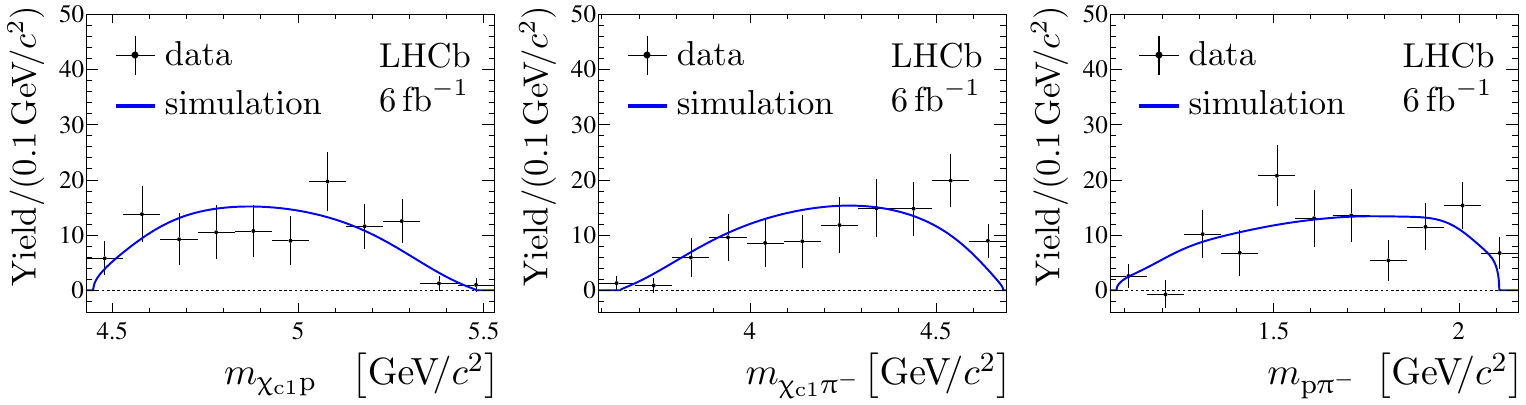} }
{  \setlength{\unitlength}{1mm}
   \begin{picture}(150,40)
		%%   
		%% 
        %% \graphpaper[5](-10,-10)(170,60)
        %%
        %% 
		\put(  0,3){\includegraphics*[width=50mm]{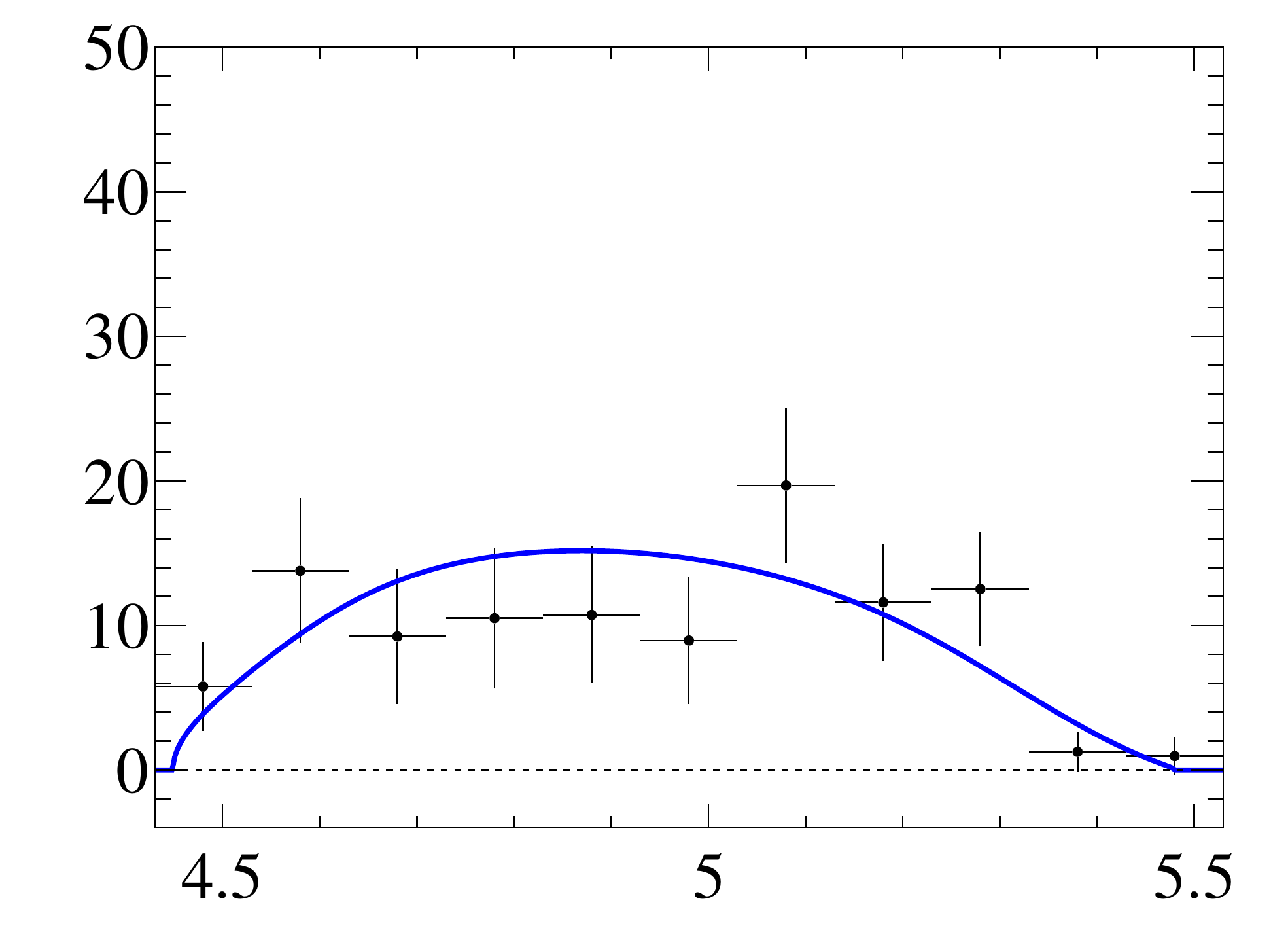}}
		\put( 52,3){\includegraphics*[width=50mm]{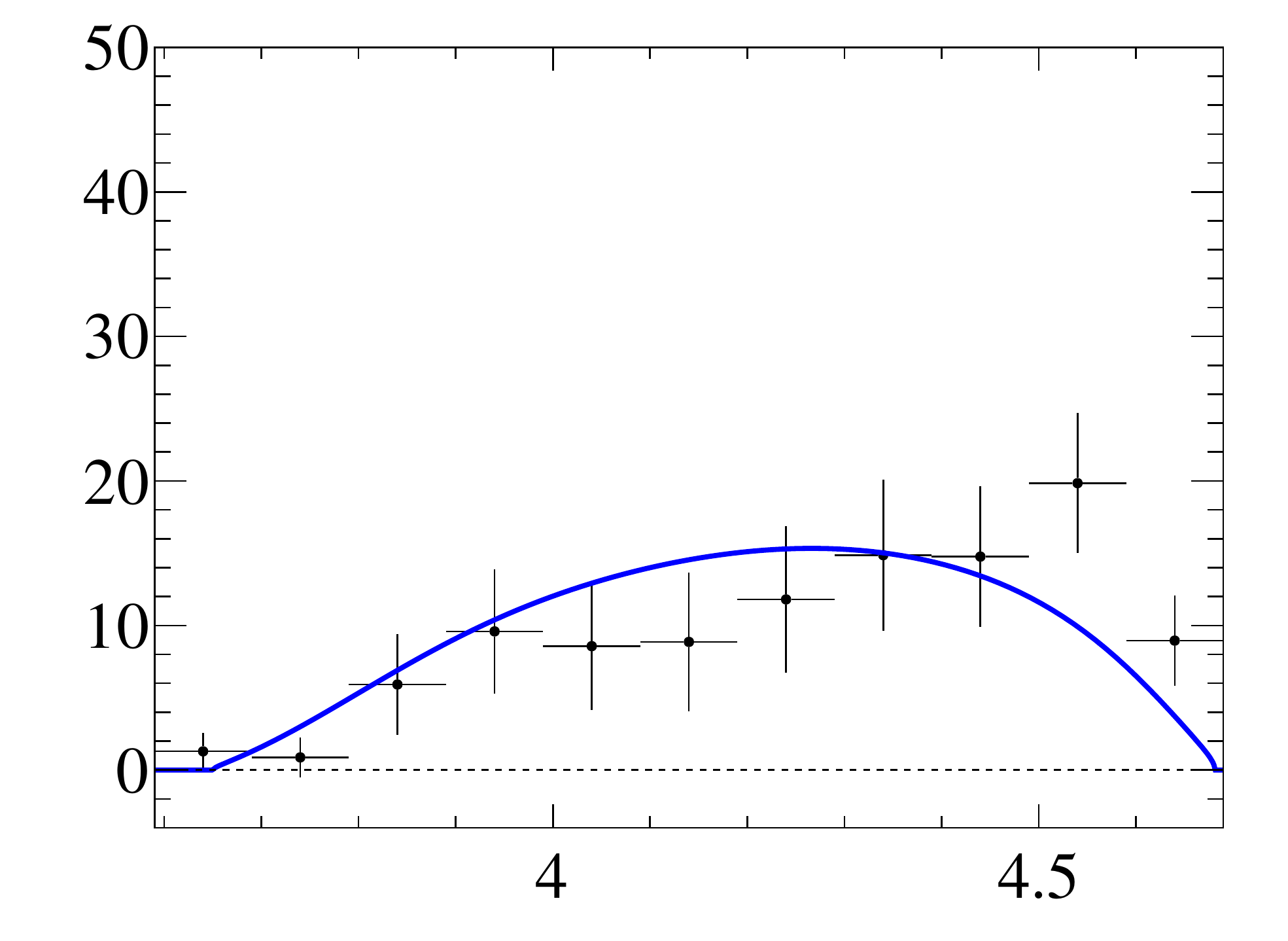}}
		\put(104,3){\includegraphics*[width=50mm]{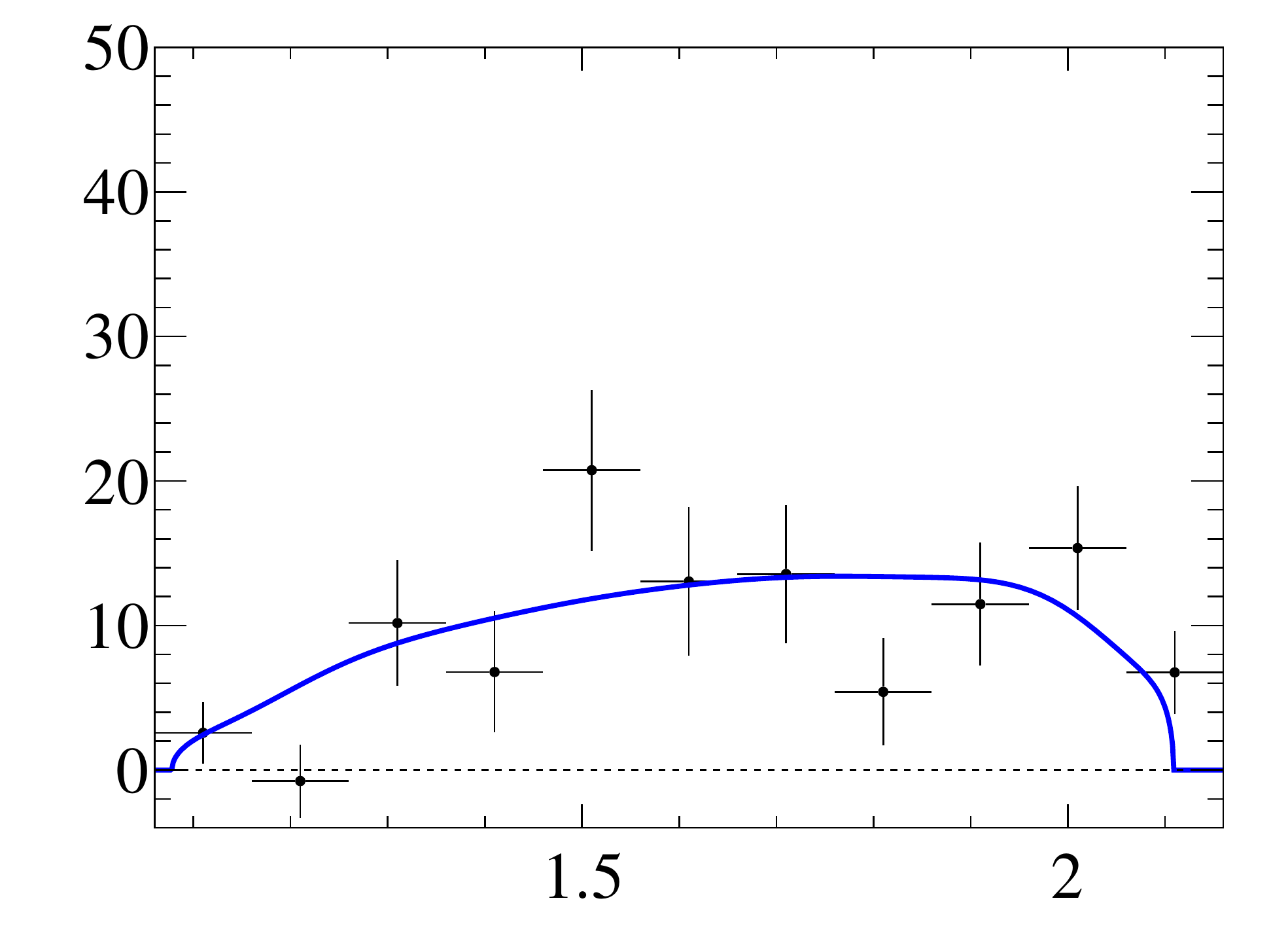}}
		\put( -2,10){\footnotesize\begin{sideways}Yield/(0.1\gevcc)\end{sideways}}
		\put( 50,10){\footnotesize\begin{sideways}Yield/(0.1\gevcc)\end{sideways}}
		\put(102,10){\footnotesize\begin{sideways}Yield/(0.1\gevcc)\end{sideways}}
		\put(20,0){\small$m_{\ChiconeP}$}
		\put(73,0){\small$m_{\ChiconePim}$ }
		\put(126,0){\small$m_{\PPim}$ }
		\put(33,0){\small$\left[\!\gevcc\right]$}
		\put(85,0){\small$\left[\!\gevcc\right]$}
		\put(137,0){\small$\left[\!\gevcc\right]$}
		\put(34,30){\footnotesize$\begin{array}{l}\lhcb\\ 6\invfb\end{array}$}
		\put(86,30){\footnotesize$\begin{array}{l}\lhcb\\ 6\invfb\end{array}$}
		\put(138,30){\footnotesize$\begin{array}{l}\lhcb\\ 6\invfb\end{array}$}
		\put(9,33.5){\line(1,0){4}}
        \put(11,31.5){\line(0,1){4}}
        \put(11,33.5){\circle*{0.53}}
        \put(61,33.5){\line(1,0){4}}
        \put(63,31.5){\line(0,1){4}}
        \put(63,33.5){\circle*{0.53}}
        \put(113,33.5){\line(1,0){4}}
        \put(115,31.5){\line(0,1){4}}
        \put(115,33.5){\circle*{0.53}}
		\put(9,28.1){\color[rgb]{0,0,1} {\rule{4mm}{1.2pt}}}
		\put(61,28.1){\color[rgb]{0,0,1} {\rule{4mm}{1.2pt}}}
		\put(113,28.1){\color[rgb]{0,0,1} {\rule{4mm}{1.2pt}}}
		\put(14,32.5){\footnotesize{data}}
		\put(66,32.5){\footnotesize{data}}
		\put(118,32.5){\footnotesize{data}}
		\put(14,27.5){\footnotesize{simulation}}
		\put(66,27.5){\footnotesize{simulation}}
		\put(118,27.5){\footnotesize{simulation}}
	\end{picture}
	}
	\caption{\small
		Background-subtracted 
		mass distributions of the~(left)~\ChiconeP, 
		(centre)~\ChiconePim~and (right)~\PPim~combinations 
		in the~\mbox{\LbChiconePPi} decay.
		Expectations from a~phase space simulation are overlaid.
	}
	\label{fig:resonances}
\end{figure}

Background-subtracted 
$\chicone\proton$,
$\chicone\pim$ and $\proton\pim$~mass distributions
from the~\mbox{\LbChiconePPi} decay
are shown in Fig.~\ref{fig:resonances}.
The~\sPlot technique,
with the $\chicone\proton\pim$~mass as 
the~discriminating variable,
is used for background subtraction~\cite{Pivk:2004ty}. 
The~distributions are compared with 
those obtained from simulated decays 
generated according 
to a~phase space model
and, with the~present dataset,  no evidence for large 
contributions from possible exotic states is found. 

%% file: systematics.tex
\section{Systematic uncertainties}

Since the \LbChicjPPi and \LbChicjPK decay 
channels have similar kinematics and topologies, systematic 
uncertainties largely cancel in the~ratios $\mathcal{R}$
defined by Eqs.~\eqref{eq:R}. 
The remaining contributions to the systematic uncertainties 
are summarized in Table~\ref{tab:systematics} and discussed below.

The~systematic uncertainty related to the~signal 
and background shapes is investigated using alternative
parameterisations. 
For the~\LbChiconePPi and \LbChiconePK components, 
two alternative models are probed.
The~first model consists of a~sum of a~Student's
$t$\nobreakdash-distribution~\cite{Student}
and a~double\nobreakdash-sided 
Crystal Ball function\,(CB$_2$) 
with power\nobreakdash-law tails 
on both sides of the~peak~\cite{LHCb-PAPER-2011-013}.
The~second alternative model is a~sum of a~Gaussian and 
CB$_2$~functions.
For the~\LbChictwoPPi and \LbChictwoPK components,  
two other  alternative models are probed:
a~sum of a~bifurcated Student's $t$\nobreakdash-distribution 
with  a~CB$_2$~function,
and a~sum of a~skewed Gaussian function~\cite{Hagan} 
with a~CB$_2$~function.  
The~alternative parameterisations 
for the~component from partially reconstructed 
\Lb~decays include
a~bifurcated Gaussian function
and a~Student's $t$-distribution.
Two~alternative shapes are used for the~background 
parameterisation.
The~first model consists of a~product of an~exponential function 
and a~second-order positive polynomial function, while 
a~fourth\nobreakdash-order  concave positive  polynomial 
function is used as the~second alternative model.
The systematic uncertainty related to the fit model is estimated 
by producing pseudoexperiments  
generated with the~baseline fit model and fitted with 
alternative models. Each pseudoexperiment is approximately 
100 times larger than the data sample. 
The~maximal deviations for the~ratios of 
the~signal yields with respect 
to the~baseline model are taken as systematic 
uncertainties in 
the~ratios~$\mathcal{R}$.
%, defined by Eqs.~\eqref{eq:R}.
The~assigned systematic uncertainties 
are 2.4\%, 3.7\% and 3.7\% in the~ratios 
\RPiK, \RTwoOnePi and \RTwoOneK, respectively. 

\begin{table}[tb]
	\centering
	\caption{
		Relative systematic uncertainties (in~\%) in 
		the~ratios of branching fractions.
		The~total uncertainty is obtained 
		as the~sum of individual components in quadrature. 
		Empty cells  correspond to 
		cases where no uncertainty is applicable.
	}
	\begin{tabular}{lccc}
		Source & $\phantom{<}\RPiK$ & \RTwoOnePi & \RTwoOneK
        \\[1.5mm]
        \hline 
        \\[-2mm]
		Fit model
		 & $\phantom{<}\systFitmodelRPiK$ & \systFitmodelRTwoOnePi & $\phantom{<}\systFitmodelRTwoOneK$ \\
		\Lb~production spectra    & $<0.1$           & &        \\
		$\LbChicjPK$~decay models & $<0.1$           & & $<0.1$ \\
		Track reconstruction      & $<0.1$           & &        \\
		Hadron identification     & $\phantom{<}0.3$ & &        \\
		Trigger efficiency 
		& $\phantom{<}\systTrigger$ & & \\
		BDTG selection 
		& $\phantom{<}\systDaSiAgr$ & & \\
		Simulation sample size 
		& $\phantom{<}\systSSSPiK$ & 
		\systSSSTwoOnePi & 
		$\phantom{<}\systSSSTwoOneK$
        \\[1.5mm]
        \hline 
        \\[-2mm]
        %%
		%% Sum in quadrature 
		Total 
		& $\phantom{<}\systPiK$ & \systTwoOnePi & $\phantom{<}\systTwoOneK$
	\end{tabular}
	\label{tab:systematics}
\end{table}

An~additional systematic uncertainty in 
the~ratios $\mathcal{R}$ arises due to
differences between data and  simulation. 
The~transverse momentum  and rapidity spectra of the~\Lb~baryons 
in simulated samples are adjusted to match those 
observed in a~high\nobreakdash-yield 
low\nobreakdash-background sample of 
reconstructed \mbox{\LbJpsiPK}~decays.
The~finite size of this sample 
%% \mbox{$\decay{\Lb}{\jpsi\proton\Km}$}~sample
causes uncertainty in the~obtained \Lb~production 
spectra.
The~systematic uncertainty in the~efficiency 
ratios, related to the~imprecise knowledge 
of the~production \Lb~baryon spectra
is estimated using the variation 
of the~kinematic spectra of 
the~selected \mbox{$\decay{\Lb}{\jpsi\proton\Km}$}~sample
within their~statistical uncertainty. 
This~systematic uncertainty is found to be 
smaller than 0.1\% in the~\RPiK~ratio and 
even smaller for the~\RTwoOnePi and \RTwoOneK~ratios.

The~simulated \LbChicjPK decays are 
corrected to reproduce the~\PKm~mass 
and \costhPKm~distributions observed 
in data. 
The~systematic uncertainty 
in the~$\upvarepsilon_{\LbChiconePK}$ 
and $\upvarepsilon_{\LbChictwoPK}$~efficiencies, related to 
the~imprecise knowledge of 
the~decay model for the~\mbox{\LbChicjPK}~decays, 
is estimated using the~variation
of the~\PKm~mass and \costhPKm~spectra 
within their uncertainties.
The~corresponding systematic uncertainties
in the~\RPiK and \RTwoOneK~ratios
are found to be less than~0.1\%. 

There are residual differences in 
the~reconstruction efficiency 
of charged\nobreakdash-particle tracks that 
do not cancel completely in the~ratio due 
to the~different kinematic distributions 
of the~final\nobreakdash-state particles.
The~track\nobreakdash-finding efficiencies 
obtained from 
simulated samples 
are corrected \mbox{using}
calibration channels~\cite{LHCb-DP-2013-002}.
%% data\nobreakdash-driven techniques
The~uncertainties related to~the~efficiency 
correction factors, 
are propagated to the~ratios of 
the~total efficiencies using pseudoexperiments
and found to be smaller than 0.1\% in 
the~ratio \RPiK and 
smaller in 
the~ratios \RTwoOnePi and \RTwoOneK.
A~small difference between data and simulation 
for the~photon reconstruction
is studied using a~large sample of 
\mbox{$\decay{\Bu}{\jpsi
\left(\decay{\Kstarp}{\Kp \left(\decay{\piz}{\g\g}\right)}
\right)}$}~decays~\cite{LHCb-PAPER-2012-022,
LHCb-PAPER-2012-053,
Govorkova:2015vqa}
%% Govorkova:2015buu
%% }. 
The~associated systematic 
uncertainty largely cancels in the~ratios~$\mathcal{R}$.

The~combined detector response used
for the~identification of protons, kaons and pions
in simulation is resampled from 
control channels~\cite{LHCb-DP-2018-001}. 
The~systematic uncertainty obtained 
through this procedure
arises from the~kernel shape  
used in the~estimation of the~probability 
density distributions. 
An~alternative combined response is estimated
using an~alternative kernel estimation 
with a changed shape and the~efficiency models 
are 
regenerated~\cite{LHCb-PAPER-2020-025,Poluektov:2014rxa}. 
The~difference between the~two estimates for 
the~efficiency ratios is taken as the~systematic 
uncertainty related to hadron identification
and is found to be 0.3\% 
in the~ratio $\RPiK$. 
In~the~ratios \RTwoOnePi and  \RTwoOneK
this systematic uncertainty 
cancels as it is assumed to be
fully correlated between the modes with 
\chicone and \chictwo~mesons.

A~systematic uncertainty 
in the~ratios related 
to the~knowledge of the~trigger efficiencies 
has been previously studied using high-yield 
\BuJpsiKp and \BuPsiKp decays by comparing 
ratios of trigger efficiencies in data 
and simulation~\cite{LHCb-PAPER-2012-010}.
Based on these comparisons, a~relative uncertainty 
of $\systTrigger\%$ is assigned to~\RPiK, 
while for~\RTwoOnePi and \RTwoOneK 
it is expected 
to cancel in the~ratio due to 
resemblance of the~kinematics
of the~corresponding decay channels.

The imperfect data description by the simulation 
due to remaining effects is studied by varying 
the BDTG selection criteria in ranges that lead 
to $\pm20\%$ changes in the measured efficiency.
For this study,  
the~high\nobreakdash-statistics normalisation 
channel is used.
The~resulting difference between 
the~efficiency estimated 
using data and simulation does 
not exceed $\systDaSiAgr\%$, 
which is taken as a systematic uncertainty in \RPiK.
This systematic uncertainty in \RTwoOnePi and \RTwoOneK 
is considered negligible due to 
the~similarity of the~kinematics 
of the~corresponding decay channels.

Finally, the~uncertainties in the~ratios 
of efficiencies from Eqs.~\eqref{eq:efficiency}
are $\systSSSPiK\%$, $\systSSSTwoOnePi\%$ and 
$\systSSSTwoOneK\%$ and are taken as 
systematic uncertainties due to 
the~finite size of the~simulated samples 
for the~\RPiK, \RTwoOnePi and \RTwoOneK, respectively.

For~each choice of the~fit model, 
the~statistical significance of 
the~\LbChictwoPPi~signal
is calculated from data 
using Wilks' theorem~\cite{Wilks:1938dza}
and confirmed by simulating a~large number 
of pseudoexperiments. 
The~smallest significance
found is $3.5$~standard deviations, 
taken as its~significance 
including~systematic uncertainties.

%% file: results.tex
\section{Results and summary}

A~search for the~Cabibbo\nobreakdash-suppressed decays 
\LbChicjPPi is performed using 
a~data sample 
collected by the~LHCb experiment 
in proton\nobreakdash-proton 
collisions at a~centre\nobreakdash-of\nobreakdash-mass 
energy of $13\tev$ and 
corresponding to 6\invfb of integrated 
luminosity.
The~$\LbChiconePPi$~decay is observed 
for the~first 
time with a~yield of \yieldLbChiconePPi
and a~statistical significance
above 9~standard deviations.
First evidence for the~\LbChictwoPPi decay is obtained 
with a~yield of \yieldLbChictwoPPi and 
a~significance of 3.5~standard deviations.
The~ratios of the~branching fractions
are measured to be
\begingroup
\allowdisplaybreaks
\begin{eqnarray*}
\RPiK = 
	\dfrac{\BR\left(\LbChiconePPi\right)}{\BR\left(\LbChiconePK\right)}
	 & = & \left( 6.59 \pm 1.01 \pm 0.22 \right)\times 10^{-2} \,, %%\valueRPiK\,,
	 \\
	 \RTwoOnePi = 
	 \dfrac{\BR\left(\LbChictwoPPi\right)}{\BR\left(\LbChiconePPi\right)}
	& = & \valueRTwoOnePi\,,
	\\
	\RTwoOneK = 
	\dfrac{\BR\left(\LbChictwoPK\right)}{\BR\left(\LbChiconePK\right)}
	& = & \valueRTwoOneK\,,
\end{eqnarray*}
\endgroup
where the~first uncertainty is statistical, 
the~second is systematic and the~third 
is related to the~uncertainties in the~branching fractions 
of the~\mbox{$\decay{\chicj}{\jpsi\g}$}~decays~\cite{PDG2020}.
The~ratio \RPiK is similar to  
analogous ratios 
for other Cabibbo\nobreakdash-suppressed 
decays of the~\Lb~baryon\cite{LHCb-PAPER-2014-020,LHCb-PAPER-2018-022}.
The~expected value for the~ratio \RPiK, 
if neglecting the~resonance structures in 
the~\LbChiconePPi and \LbChiconePK~decays, is 
%% the~expected value for the~ratio \RPiK is
%
\begin{equation*}
%% \mathcal{R}_{\pion/\kaon}^{\mathrm{th}}
%% \approx 
  \dfrac{ \Phi_3\left(\LbChiconePPi\right)}
  { \Phi_3\left(\LbChiconePK\right)} \times
  \tan^2 \uptheta_{\mathrm{C}} \simeq 9.9\%\,,
\end{equation*}
where $\Phi_3$ denotes the~full 
three\nobreakdash-body phase space and 
$\uptheta_{\mathrm{C}}$ is the~Cabibbo angle~\cite{Cabibbo:1963yz}.
The~ratio \RTwoOneK 
agrees well with the~previous 
measurement by the~LHCb collaboration 
of \valueRTwoOneKPrevious~\cite{LHCb-PAPER-2017-011}.
This result has better precision and arises 
from a~statistically independent sample 
from that of Ref.~\cite{LHCb-PAPER-2017-011}.
Similarly to \RTwoOneK, 
the~new result for \RTwoOnePi shows 
no suppression 
of the~\chictwo mode relative to the~\chicone~mode, 
which challenges 
the~factorisation approach 
for \Lb~decays~\cite{Beneke:2008pi}.  

The background\nobreakdash-subtracted
\ChiconeP and \ChiconePim mass 
distributions for the \LbChiconePPi decay 
are investigated.
With the~present dataset, the~results are consistent 
with a~phase space model, 
and no evidence for contributions from exotic states is found.
%% and, with 
%% the~present %%% low statistics 
%% dataset, no evidence for contributions from 
%% exotic states is found.

%% file: acknowledgements.tex
\section*{Acknowledgements}
%
% These Acknowledgements valid from 3-May-2019
%
\noindent We express our gratitude to our colleagues 
in the~CERN
accelerator departments for 
the~excellent performance of the~LHC. 
We~thank the~technical and administrative staff 
at the~LHCb institutes.
We~acknowledge support from CERN and from the national agencies:
CAPES, CNPq, FAPERJ and FINEP\,(Brazil); 
MOST and NSFC\,(China); 
CNRS/IN2P3\,(France); 
BMBF, DFG and MPG\,(Germany); 
INFN\,(Italy); 
NWO\,(Netherlands); 
MNiSW and NCN\,(Poland); 
MEN/IFA\,(Romania); 
MSHE\,(Russia); 
MICINN\,(Spain); 
SNSF and SER\,(Switzerland); 
NASU\,(Ukraine); 
STFC\,(United Kingdom); 
DOE NP and NSF\,(USA).
We~acknowledge the~computing resources that are provided by CERN, 
IN2P3\,(France), 
KIT and DESY\,(Germany), 
INFN\,(Italy), 
SURF\,(Netherlands),
PIC\,(Spain), 
GridPP\,(United Kingdom), 
RRCKI and Yandex LLC\,(Russia), 
CSCS\,(Switzerland), 
IFIN\nobreakdash-HH\,(Romania), 
CBPF\,(Brazil),
PL\nobreakdash-GRID\,(Poland) 
and NERSC\,(USA).
We~are indebted to the~communities behind 
the~multiple open\nobreakdash-source
software packages on which we depend.
Individual groups or members have received support from
ARC and ARDC\,(Australia);
AvH Foundation\,(Germany);
EPLANET, Marie Sk\l{}odowska\nobreakdash-Curie Actions and ERC\,(European Union);
A*MIDEX, ANR, Labex P2IO and OCEVU, and 
R\'{e}gion Auvergne\nobreakdash-Rh\^{o}ne\nobreakdash-Alpes\,(France);
Key Research Program of Frontier Sciences of CAS, CAS PIFI, CAS CCEPP, 
Fundamental Research Funds for the~Central Universities, 
and Sci. \& Tech. Program of Guangzhou\,(China);
%Key Research Program of Frontier Sciences of CAS, CAS PIFI,
%Thousand Talents Program, and Sci. \& Tech. Program of Guangzhou (China);
RFBR, RSF and Yandex LLC\,(Russia);
GVA, XuntaGal and GENCAT\,(Spain);
the~Leverhulme Trust, the~Royal Society
 and UKRI\,(United Kingdom).

%% file: Authorship_LHCb-PAPER-2021-003.tex
% LHCb collaboration author list
% Data extracted on January 28th, 2021 at 11:37am for paper reference LHCb-PAPER-2021-003
\centerline
{\large\bf LHCb collaboration}
\begin
{flushleft}
\small
R.~Aaij$^{32}$,
C.~Abell{\'a}n~Beteta$^{50}$,
T.~Ackernley$^{60}$,
B.~Adeva$^{46}$,
M.~Adinolfi$^{54}$,
H.~Afsharnia$^{9}$,
C.A.~Aidala$^{85}$,
S.~Aiola$^{25}$,
Z.~Ajaltouni$^{9}$,
S.~Akar$^{65}$
% ...

{\footnotesize \it

% ...
$^{9}$Universit{\'e} Clermont Auvergne, CNRS/IN2P3, LPC, Clermont-Ferrand, France\\
% ...
$^{25}$INFN Sezione di Milano, Milano, Italy\\
% ...
$^{32}$Nikhef National Institute for Subatomic Physics, Amsterdam, Netherlands\\
% ...
$^{46}$Instituto Galego de F{\'\i}sica de Altas Enerx{\'\i}as (IGFAE), Universidade de Santiago de Compostela, Santiago de Compostela, Spain\\
% ...
$^{50}$Physik-Institut, Universit{\"a}t Z{\"u}rich, Z{\"u}rich, Switzerland\\
% ...
$^{54}$H.H. Wills Physics Laboratory, University of Bristol, Bristol, United Kingdom\\
% ...
$^{60}$Oliver Lodge Laboratory, University of Liverpool, Liverpool, United Kingdom\\
% ...
$^{65}$University of Cincinnati, Cincinnati, OH, United States\\
% ...
$^{68}$Syracuse University, Syracuse, NY, United States\\
% ...
$^{85}$University of Michigan, Ann Arbor, United States, associated to $^{68}$\\
\bigskip
% ...
\medskip
}
\end{flushleft}